\documentclass[prb,amsmath,amssymb,twocolumn]{revtex4-2}
\usepackage{graphicx}
\usepackage{subfigure}
\usepackage{adjustbox}
\usepackage{bm}
\usepackage{color}
\usepackage{braket}
\usepackage{standalone}
\usepackage{multirow}
\usepackage{tikz}
\usepackage{mathrsfs}
\usepackage{dsfont}
\usepackage{comment}
\usepackage{amsmath}
\usepackage[colorlinks,bookmarks=true,citecolor=blue,linkcolor=red,urlcolor=blue]{hyperref}
\usepackage[capitalize]{cleveref}

\newcommand{\be}{\begin{equation}}
\newcommand{\ee}{\end{equation}}

\newcommand{\ba}{\begin{eqnarray}}
\newcommand{\ea}{\end{eqnarray}}

\newcommand*{\id}{{\normalfont\hbox{1\kern-0.15em \vrule width .8pt depth-.5pt}}}

\newcommand{\commentblock}[1]{}

\graphicspath{{./figures/}}

\begin{document}

\title{Deformed Fredkin model for the $\nu{=}5/2$ Moore-Read state on thin cylinders}

\author{Cristian Voinea$^1$, Songyang Pu$^1$, Ammar Kirmani$^2$, Pouyan Ghaemi$^{3,4}$, Armin Rahmani$^5$, and Zlatko Papi\'{c}$^1$}

\affiliation{${}^1$School of Physics and Astronomy, University of Leeds, Leeds LS2 9JT, United Kingdom}
\affiliation{${}^2$Theoretical Division, T-4, Los Alamos National Laboratory, Los Alamos, NM 87545, USA}
\affiliation{${}^3$Physics Department, City College of the City University of New York, NY 10031, USA}
\affiliation{${}^4$Physics Program, Graduate Center of City University of New York, NY 10031, USA}
\affiliation{${}^5$Department of Physics and Astronomy and Advanced Materials Science and Engineering Center, Western Washington University, Bellingham, Washington 98225, USA}

\date{\today}

\begin{abstract}
We propose a frustration-free model for the Moore-Read quantum Hall state on sufficiently thin cylinders with circumferences $\lesssim 7$ magnetic lengths. While the Moore-Read Hamiltonian involves complicated long-range interactions between triplets of electrons in a Landau level, our effective model is a simpler one-dimensional chain of qubits with deformed Fredkin gates. We show that the ground state of the Fredkin model has high overlap with the Moore-Read wave function and accurately reproduces the latter's entanglement properties.  Moreover, we demonstrate that the model captures the dynamical response of the Moore-Read state to a geometric quench, induced by suddenly changing the anisotropy of the system. We elucidate the underlying mechanism of the quench dynamics and show that it coincides with the linearized bimetric field theory. The minimal model introduced here can be directly implemented as a first step towards quantum simulation of the Moore-Read state, as we demonstrate by deriving an efficient circuit approximation to the ground state and implementing it on IBM quantum processor.

\end{abstract}

\maketitle

\section{Introduction}\label{sec:intro}

The enigmatic fractional quantum Hall (FQH) state observed at filling fraction $\nu{=}5/2$~\cite{Willett87} stands out as a rare example of an even-denominator state among the majority of odd-denominator states  described by the Laughlin wave functions~\cite{Laughlin83} and composite fermion theory~\cite{Jain89}. One of the leading theoretical explanations of the $\nu{=}5/2$ state is based on the Moore-Read (MR) variational wave function~\cite{Moore91}. 
Two unique properties of the MR state are worth highlighting: (i) it represents a $p$-wave superconductor of composite fermions~\cite{Read00}; (ii) its elementary charge excitations behave like Ising anyons, i.e., they carry charge $e/4$ and exhibit non-Abelian braiding statistics~\cite{Moore91,Nayak96}. The latter has motivated the use of MR state as a potential resource for topological quantum computation~\cite{Nayak08}, whereby quantum information is encoded in the collective states of MR anyons and quantum gates are executed by braiding the anyons. Such operations would be protected by the topological FQH gap, avoiding the costly quantum error correction. 

On the fundamental side, the understanding of particle-hole symmetry and collective excitations in the $\nu{=}5/2$ state has recently generated a flurry of interest. While the  numerics~\cite{Morf98, Rezayi00} provided initial support of the MR wave function capturing the physical ground state at $\nu{=}5/2$, it has been realized that preserving (or breaking) particle-hole (PH) symmetry can lead to distinct phases of matter. For example, by PH-conjugating the MR wave function one obtains a distinct state known as the ``anti-Pfaffian'' state~\cite{Levin07, Lee07}, while enforcing the PH symmetry leads to another, PH-symmetric Pfaffian state (``PH-Pf'')~\cite{Son15}. Understanding the relation of these states with the MR state in light of physical PH symmetry breaking effects, such as Landau level mixing~\cite{Bishara09, Sodemann13, Simon13} remains an important task for reconciling numerics~\cite{Pakrouski15,Rezayi17,Balram18, Mishmash18} with experiment~\cite{Banerjee18, Dutta21}. 

On the other hand, collective excitations of the $\nu{=}5/2$ state have also attracted much attention. The pairing in the MR state mentioned above gives rise to an additional collective mode -- the unpaired ``neutral fermion" mode -- which has been ``seen'' in the numerical simulations~\cite{Bonderson11NF, Moller11, Papic12}, but so far not detected in experiment. The gap of the neutral fermion mode is of direct importance for topological quantum computation, as the former can be excited in the process of fusion of two elementary anyons. Recently, Ref.~\cite{Gromov20} proposed a description of the neutral fermion mode based on an emergent ``supersymmetry'' with the more conventional, bosonic density-wave excitation~\cite{Girvin85,Girvin86}. The numerics in Ref.~~\cite{Pu23} suggests that supersymmetry can indeed emerge in a realistic microscopic model of $\nu{=}5/2$. 

\begin{figure}[b]
    \centering
    \includegraphics[width=\linewidth]{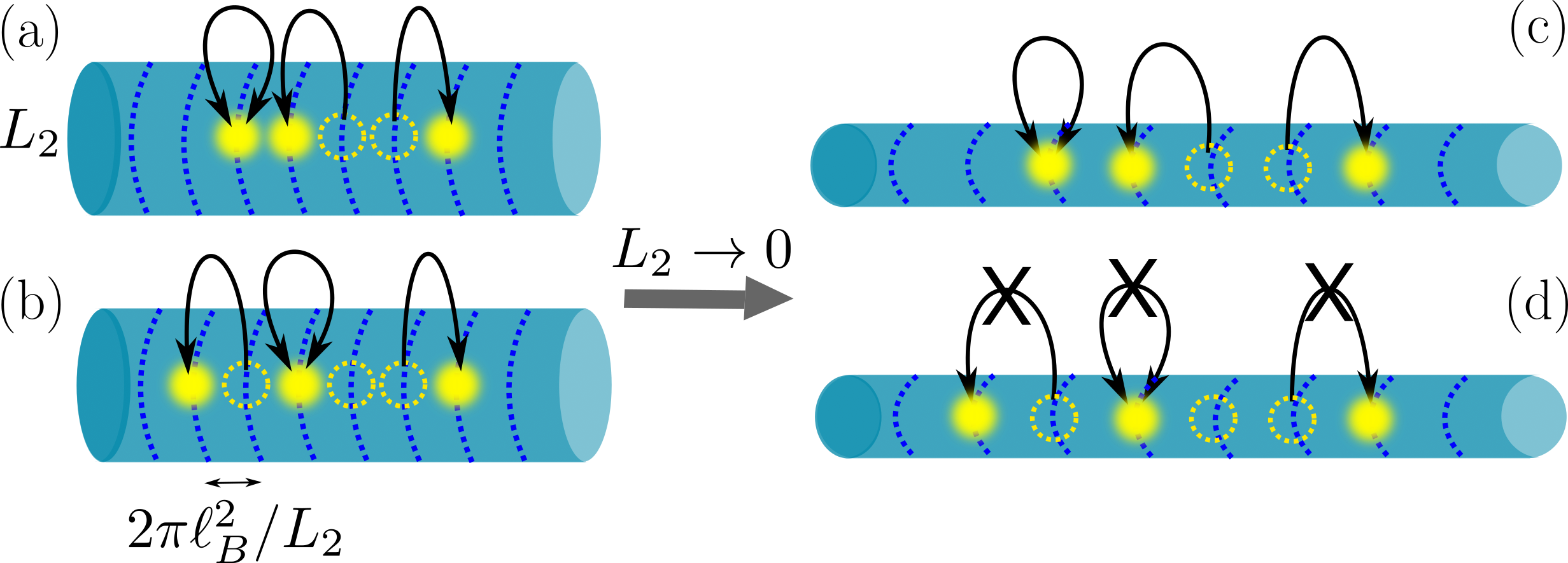}
    \caption{(a)-(b): Two types of 3-electron scattering processes present in the Moore-Read Hamiltonian. The cylinder circumference, $L_2$, controls the spacing $2\pi\ell_B^2/L_2$ between Landau level orbitals (dashed lines). (c)-(d): Sending $L_2/\ell_B{\to}0$ suppresses the longer-range hopping (d) compared to the one in (c). It will be shown that (c), where one electron is fixed while the other two electrons hop between the nearest-neighbor orbitals, maps to a controlled-SWAP (Fredkin) gate.   }
    \label{fig:hopping}
\end{figure}

In this paper we develop a framework for studying the MR state in a quasi-one-dimensional limit, obtained by placing the FQH fluid on a streched cylinder or torus whose lengths in the two directions obey $L_2 {\ll} L_1$.  This so-called  ``thin-torus'' limit has been fruitful in gaining understanding of the structure of many FQH ground states and their excitations~\cite{Tao83,Rezayi94b, Seidel05,Bergholtz05,Bergholtz08b,Seidel08,Seidel11,Weerasinghe14}. The thin-torus limit provides a natural classical ``cartoon'' of the complicated physics in the two-dimensional limit: the off-diagonal matrix elements of the Hamiltonian describing a FQH state become strongly suppressed ${\sim}\exp[-(2\pi\ell_B/L_2)^2]$ in the limit $L_2/\ell_B{\to}0$, allowing for considerable simplications of the problem --  see Fig.~\ref{fig:hopping} for an illustration.  

However, there have been comparatively few studies of the MR state near the thin-torus limit. Most previous works~\cite{Seidel06, Bergholtz06Pf, Seidel08Pf, Flavin11} focused on the ``extreme'' thin-torus limit, also known as the Tao-Thouless limit~\cite{Tao83}, where the Hamiltonian is reduced to purely classical electrostatic repulsion. It is therefore important to develop an analytically-tractable model for the MR state \emph{beyond} the strict Tao-Thouless limit, where some correlated hopping terms are present in the model. For the Laughlin and Jain states, such models were previously formulated in Refs.~\cite{Nakamura2012,Wang12f,Nakamura11,Soule12} and one of the goals of this paper is to work out an analogous model for the MR state.  The intrinsic one-dimensional (1D) structure of such models makes them suitable for implementation on digital quantum computers, as recently shown for the $\nu{=}1/3$ Laughlin state~\cite{Rahmani20,Kirmani2021}. The versatility of such devices allows to probe questions, such as the real-time dynamics following a quench,  that are challenging for traditional solid-state experiments~\cite{Pinczuk93, Platzman96, Kang01, Kukushkin09}. In particular, the implementation on IBM quantum processor allowed to simulate the ``graviton" dynamics induced by deforming the geometry of the Laughlin state~\cite{Haldane11, Liu18, Lapa19, Liou19, Nguyen2021}. 

The remainder of this paper is organized as follows. We start by reviewing the parent Hamiltonian of the MR state in Sec.~\ref{sec:MRHamiltonian}. We make use of the second-quantization formalism to derive a simplified frustration-free model near the thin-cylinder limit and we show that its ground state has high overlap with the MR state, with similar entanglement properties. In Sec.~\ref{sec:spinmodel} we show that the frustration-free model can be expressed
as a deformed Fredkin model for spin-1/2 degrees of freedom. Working in the spin representation, we present an intuitive picture of the ground state of this deformed Fredkin model and derive its matrix-product state (MPS) representation. We also demonstrate that the ground state can be efficiently approximated by a quantum circuit, which we implement on the IBM quantum processor. In Sec.~\ref{sec:dynamics} we show that the Fredkin model also captures the dynamics of the MR state induced by quenching the anisotropy of the system, and we elucidate the mechanism of this dynamics. Our conclusions are presented in Sec.~\ref{sec:conclusions}, while Appendices contain technical details of the derivations, further characterizations of the ground state, and a generalization of the Laughlin case in Ref.~\cite{Nakamura2012} to a closely-related Motzkin spin chain.

\section{Moore-Read Hamiltonian on a thin cylinder} \label{sec:MRHamiltonian}

In this section we formulate a frustration-free model that provides a good approximation of the MR ground state near the thin-cylinder limit. In the infinite 2D plane, the parent Hamiltonian of the MR state is a peculiar interaction potential that penalizes configurations of any three electrons forming a state with relative angular momentum equal to 3 -- the smallest possible momentum allowed by the Pauli exclusion principle~\cite{Greiter92a,Read96}. At the same time, pairs of electrons do not experience any interaction. The combination of these two effects gives rise to an exotic many-electron state with $p$-wave pairing correlations~\cite{Read00}.  

Concretely, the MR interaction potential can be written in real space using derivatives of delta functions~\cite{Rezayi00}: 
\begin{equation}\label{eq:MRHam}
H_\mathrm{MR} =-\sum_{i<j<k} S_{i, j, k}\left[\nabla_i^4 \nabla_j^2 \delta^2\left(\mathbf{r}_i-\mathbf{r}_j\right) \delta^2\left(\mathbf{r}_j-\mathbf{r}_k\right)\right],    
\end{equation}
where $S_{i,j,k}$ is a symmetrizer over the electron indices $i$, $j$, $k$. At filling $\nu{=}1/2$, the ground state of this Hamiltonian has energy $E{=}0$ and it is unique (on a disk, sphere or cylinder geometry) or six-fold degenerate on a torus, corresponding exactly to the wave function written down by Moore and Read~\cite{Moore91}. The same state was shown to have high overlap with the exact ground state of Coulomb interaction in the first-excited Landau level~\cite{Rezayi00,Storni10}. 
Moreover, the Hamiltonian above also captures the collective excitations of the MR state~\cite{Bonderson11NF,Moller11,Sreejith11,Papic12,Repellin15}. Below we first convert the Hamiltonian (\ref{eq:MRHam}) into a second-quantized form on the cylinder and torus geometries. This form allows us to derive a simplified model for the MR state on a thin cylinder.

\subsection{Moore-Read Hamiltonian in second quantization}\label{sec:secondquantized}

The singularities in Eq.~(\ref{eq:MRHam}) are naturally regularized by projection to the lowest Landau level (LLL). Assuming Landau gauge $\mathbf{A}=(0,Bx,0)$, the single-electron wave functions are given by~\cite{ChakrabortyBook} 
\begin{eqnarray}\label{eq:orbitals}
    \phi_j(\mathbf{r})= \frac{1}{\sqrt{L_2\sqrt{\pi}\ell_B}} e^{i 2\pi j y\ell_B^2/L_2} e^{-(x-2\pi j \ell_B^2/L_2)^2/2\ell_B^2},
\end{eqnarray}
where $L_2$ is the cylinder circumference in the $y$-direction and $\ell_B=\sqrt{\hbar c/eB}$ is the magnetic length. The $j$th magnetic orbital is therefore exponentially localized (in $x$-direction) around $2\pi j \ell_B^2/L_2$. For simplicity, unless specified otherwise, below we will work in units $\ell_B{=}1$.

The second-quantized representation of the MR Hamiltonian is:
\begin{equation} \label{eq: MR parent hamiltonian}
H_\mathrm{MR} = \sum_{j_1,\dots ,j_6=0}^{N_\phi-1} V_{j_1j_2j_3j_4j_5j_6} \, c_{j_1}^\dagger c_{j_2}^\dagger c_{j_3}^\dagger c_{j_4} c_{j_5}c_{j_6}, 
\end{equation}
where the operators $c_j^\dagger$, $c_j$ create or destroy an electron in the  orbital $\phi_j(\mathbf{r})$.  The matrix elements are derived by integrating Eq.~(\ref{eq:MRHam}) between the single-electron eigenfunctions (\ref{eq:orbitals}), which yields
\begin{align}\label{eq:matrix elements}
\nonumber V_{j_1j_2j_3j_4j_5j_6} &= \delta_{j_1+j_2+j_3,j_4+j_5+j_6} \\ 
\nonumber &\times (j_1-j_2)(j_1-j_3)(j_2-j_3)\\
\nonumber &\times (j_6-j_5)(j_6-j_4)(j_5-j_4) \\
 &\times  \exp \bigg[ -\frac{\kappa^2 }{2g_{11}} \bigg( \sum j_i^2 -\frac{1}{6}\big(\sum j_i\big)^2 \nonumber \\
 &+ ig_{12}\big(j_{1}^{2}+j_{2}^{2}+j_{3}^{2}-j_{4}^{2}-j_{5}^{2}-j_{6}^{2}\big) \bigg) \bigg],
\end{align}
where we have dropped the overall normalization constant for simplicity. The magnitude of the matrix element is controlled by the cylinder circumference $L_{2}$ in units of magnetic length $\ell_B$, which defines the parameter $\kappa {=}2\pi\ell_B/L_{2}$. We have derived the matrix elements by assuming a general anisotropic band-mass tensor $g_{ab}$, which will be relevant for the discussion of geometric quench in Sec.~\ref{sec:dynamics}. Note that the matrix element $V_{j_1\cdots j_6}$ is properly antisymmetric, resulting in a minus sign when any two electrons are exchanged, hence the limits in the sum in Eq.~(\ref{eq: MR parent hamiltonian}) can be restricted to $j_1{>}j_2{>}j_3$, $j_6{>}j_5{>}j_4$ without loss of generality. The delta function in Eq.~(\ref{eq:matrix elements}) encodes momentum conservation during a scattering process, hence one of the indices, e.g., $j_6$ can be eliminated in terms of $j_1,\ldots,j_5$.

A few comments are in order. We have denoted by integer $N_\phi$ the number of magnetic orbitals. For a cylinder, $N_\phi = 2N_e{-}2$, where $N_e$ is the number of electrons. The offset $-2$ is a geometric feature of the MR state called the Wen-Zee shift~\cite{WenZee}.  The total area of the fluid of dimensions $L_1\times L_2$ must be quantized in any FQH state~\cite{Prange87}, thus we take the thin-cylinder limit according to
\begin{eqnarray}\label{eq:thincylinderlimit}
    L_2/\ell_B\to 0, \;\;\; \text{such that} \;\;\; L_1 L_2 = 2\pi\ell_B^2 N_\phi,
\end{eqnarray}
which ensures that the number of orbitals, and hence the filling factor, remains constant.  

Although we will focus on cylinder geometry in this paper, we note that the same Hamiltonian, Eqs.(\ref{eq: MR parent hamiltonian})-(\ref{eq:matrix elements}), can also be used on a torus, with a few modifications. On a torus, the shift vanishes and $N_\phi=2N_e$. However, because of the periodicity in both $x$ and $y$ directions, the momentum is only defined modulo $N_\phi$. This means that the momentum conservation takes the form $j_1+j_2+j_3=j_4+j_5+j_6 \; (\mathrm{mod} \; N_\phi$). Moreover, the matrix element (\ref{eq:matrix elements}) must be explicitly periodized to make it compatible with the torus boundary condition, which can be done by replacing $j_i {\to} j_i {+} k_i N_\phi$ and summing over $k_i$.

The derivation of the effective Hamiltonian in the thin-cylinder limit proceeds by noting that, in the limit of $\kappa{\gg}1$ (equivalently, $L_2{\ll}\ell_B$), there is a natural hierarchy of matrix elements (\ref{eq:matrix elements}), which are separated by different powers of $\exp(-\kappa^2)$~\cite{Seidel05,Bergholtz08b,Papic14}. Below we list the first few relevant terms in decreasing order:
{\small 
\begin{align}
    &2^{2}e^{-2\kappa^2} n_{p+1}n_p n_{p-1}; \quad 111 \label{eq:mr term 1} \\
    & 2^{2}3^{2}e^{-14\kappa^2/3} n_{p+2}n_{p+1}n_{p-1}; \quad 1011 \label{eq:mr term 2}\\
    &2^{5}e^{-5\kappa^2} c^{\dagger}_{p-1}c^{\dagger}_{p}c^{\dagger}_{p+1}c_{p+2}c_{p}c_{p-2}; \quad 10101 \to 01110 \label{eq:mr term 3}\\
    &2^{3}3^{2}e^{-20\kappa^2/3} c^{\dagger}_{p}c^{\dagger}_{p+1}c^{\dagger}_{p+4}c_{p+3}c_{p+2}c_{p}; \quad 11001 \to 10110 \label{eq:mr term 4}\\
    &2^{8}e^{-8\kappa^2} n_{p+2}n_{p}n_{p-2}; \quad 10101 \label{eq:mr term 5}\\
    &2^{3}5 e^{-8\kappa^2} c^{\dagger}_{p-2}c^{\dagger}_{p+2}c^{\dagger}_{p+3}c_{p+2}c_{p+1}c_{p}; \quad 001110 \to 100011 \label{eq:mr term 6}\\
    &2^{4}3^{2}e^{-26\kappa^2/3} n_{p+4}n_{p+3}n_{p}; \quad 11001 \label{eq:mr term 7}\\
    &2^{2}3^{2}5e^{-26\kappa^2/3} c^{\dagger}_{p-1}c^{\dagger}_{p}c^{\dagger}_{p+2}c_{p+3}c_{p}c_{p-2}; \quad 101001 \to 011010. \label{eq:mr term 8}
\end{align}}
We have included a binary mnemonic to represent the type of process generated by each Hamiltonian term. A single pattern, e.g., 1011, represents a diagonal term in the Hamiltonian which assigns energy penalty for the given local pattern of occupation numbers anywhere in the system. The terms containing an arrow, such as $10011 {\to} 01101$, can be visualized as correlated hopping processes,  Fig.~\ref{fig:hopping}. In such cases, the Hermitian conjugates of the processes, corresponding to reflected hoppings with the same amplitude, are implied.

\subsection{Tao-Thouless limit}\label{sec:extreme}

The ``extreme'' thin-torus limit, also known as the Tao-Thouless limit, of the MR Hamiltonian was discussed in Refs.~\cite{Bergholtz06Pf,Seidel06}. In this limit, the only terms that survive are \cref{eq:mr term 1,eq:mr term 2}, giving energy penalty to configurations $\dots111\dots$ and $\dots1011\dots$. Hence, the ground states at filling $\nu{=}1/2$ (with zero energy) are 
\begin{eqnarray}\label{eq:extreme}
\dots 110011001100\dots \;\; \mathrm{and} \; \; \dots 10101010\dots,  
\end{eqnarray}
while any other Fock state will be higher in energy by at least an amount ${\sim}\exp(-14\kappa^2/3)$, see \cref{eq:mr term 2}.  This gives the expected 6-fold ground state degeneracy of the MR state on the torus~\cite{Read00}, since the first state in Eq.~(\ref{eq:extreme}) is 4-fold and the second is 2-fold degenerate under translations. These ground states have different momenta on the torus, so they live in different sectors of the Hilbert space~\cite{Haldane85b}. 
Similarly, the ground states in Eq.~(\ref{eq:extreme}) are the \emph{densest} zero-energy states one can construct, as increasing the filling factor would necessarily violate the terms in Eq.~(\ref{eq:mr term 1})-(\ref{eq:mr term 2}). On the other hand, \emph{decreasing} the filling factor is allowed, leading to many more $E{=}0$ states. These correspond to quasihole excitations and can be interpreted as domain walls between two different types of ground state patterns in Eq.~(\ref{eq:extreme}), see Refs.~\cite{Bergholtz06Pf,Seidel06}.

On a finite cylinder, the densest zero-energy ground state is found instead at $N_\phi{=}2N{-}2$, as expected from the Wen-Zee shift. This coincides with the root partition of the Jack polynomial corresponding to the MR state~\cite{Bernevig08}, $11001100\dots 0011$, with $11$ at each boundary. The other torus root state can be similarly adapted to a finite cylinder according to $1010\ldots 101$. However, this requires an extra orbital, since the flux is now $N_\phi{=}2N{-}1$. Thus, the second type of torus ground state becomes an excited state on a cylinder.  In both cylinder and torus geometries, the Tao-Thouless ground states are trivial product states without any entanglement. We next discuss how to go beyond the extreme thin-cylinder limit and generate an entangled ground state.

\subsection{Frustration-free model beyond the Tao-Thouless limit}

Beyond the extreme limit discussed above, we would like to retain a few more terms, with smaller powers of $\exp(-\kappa^2)$, and thereby generate a more accurate approximation of the MR state over a slightly large range of $L_2$. A natural way to do to this would be to choose a magnitude cut-off and keep only the Hamiltonian matrix elements that are larger than this cutoff. However, we would also like to be able to analytically solve for the ground state of the resulting truncated Hamiltonian. In this sense, it is natural to look for a Hamiltonian which is frustration-free, i.e., has a ground state that is simultaneously annihilated by all individual terms in the Hamiltonian. In such cases it is often possible to find analytically exact ground states even though the Hamiltonian overall may not be solvable, e.g., as in the case of the Affleck-Kennedy-Lieb-Tasaki (AKLT) model~\cite{Affleck88}. 

Unfortunately, the program outlined above fails for our 3-body Hamiltonian: keeping the terms in the order they are listed in Eqs.~(\ref{eq:mr term 1})-(\ref{eq:mr term 8}) does not result in a positive semi-definite operator. This can be seen by considering the first two correlated hoppings, \cref{eq:mr term 3,eq:mr term 4}. One would want to include these hoppings as they naturally act on the two types of ground states in the extreme thin-torus limit, Eq.~(\ref{eq:extreme}), and would create some entanglement.  However, the ``dressed'' ground states would no longer be zero modes and their degeneracy would be lifted.
Inspired by the Laughlin construction~\cite{Nakamura2012}, one could try to remedy this by including the terms \cref{eq:mr term 5,eq:mr term 7} to create a sum of \emph{two} positive semi-definite operators. One quickly realizes that the hopping term \cref{eq:mr term 8} now becomes a problem, spoiling the frustration-free property. In Ref.~\cite{Papic14}, an attempt was made to define a frustration-free model for a bosonic MR state by dropping the equivalent of hopping \cref{eq:mr term 8} (as well as the hopping \cref{eq:mr term 6}). Unfortunately, upon further inspection, we have found the claim in Ref.~\cite{Papic14} to be inaccurate because the model proposed there does not yield strictly zero-energy ground states.  

\begin{figure}[tb] 
    \includegraphics[width=1\columnwidth]{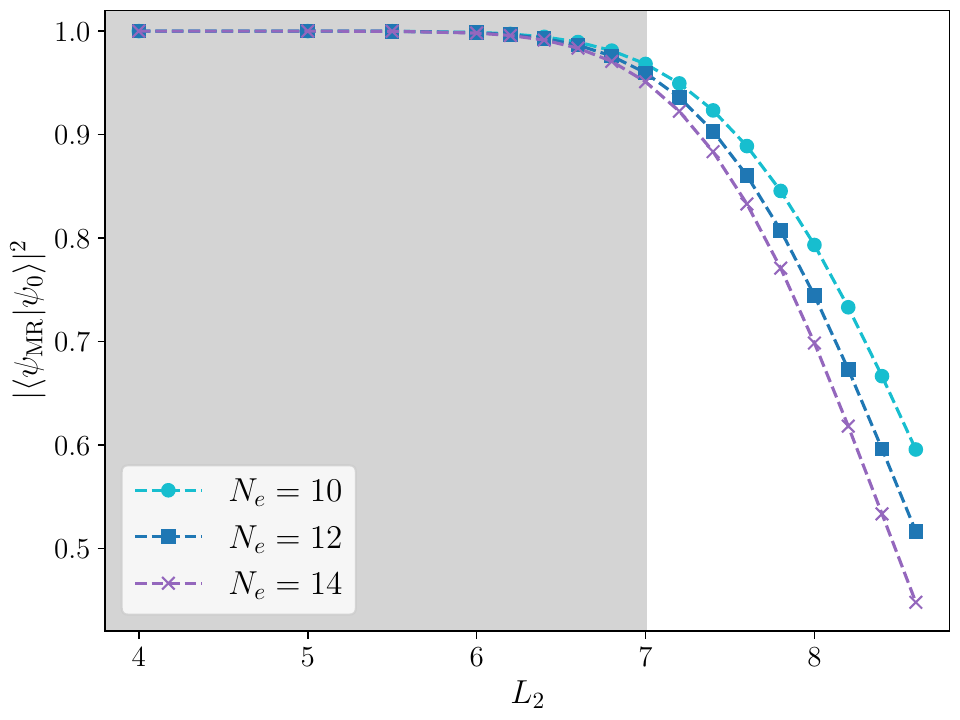}
	\caption{Overlaps between the ground state $|\psi_{\mathrm{MR}}\rangle$ of the full model in \cref{eq: MR parent hamiltonian} (i.e., the Moore-Read state) and the ground state $|\psi_{0}\rangle$ of the truncated model in \cref{eq:cylinder model}, for different system sizes. For cylinder circumferences up to $L_{2}{\approx} 7\ell_B$ (shaded), the overlap is $95\%$ or higher, indicating this truncation returns a good approximation for the ground state in the thin-cylinder regime.}
	\label{FredkinOverlap}
\end{figure}

We now describe the simplest frustration-free truncation of the Hamiltonian in Eqs.~(\ref{eq:mr term 1})-(\ref{eq:mr term 8}) that we have found. We will focus on the cylinder root state $|R_{0}\rangle = |11001100\dots0011\rangle$, which is nondegenerate.  In order to obtain a unique ``dressed'' ground state on the cylinder, we consider the Hamiltonian terms that act nontrivially on this root state. The resulting states will be the first relevant corrections to the ground state. The effective Hamiltonian contains terms in \cref{eq:mr term 1,eq:mr term 2,eq:mr term 4,eq:mr term 7}:
\begin{equation} \label{eq:cylinder model}
H_\mathrm{MR}^\prime = \sum_{i=0}^{N_{\phi}-3} A^{\dag}_{i}A_{i} + \sum_{i=0}^{N_{\phi}-4} \big(B^{\dag}_{i}B_{i} + C^{\dag}_{i}C_{i}\big),
\end{equation}
where the operators $A$, $B$ and $C$ are given by
\begin{eqnarray}
A_{i} &=& \alpha c_{i} c_{i+1} c_{i+2}, \\    
B_{i} &=& \beta c_{i} c_{i+2} c_{i+3} + \gamma c_{i} c_{i+1} c_{i+4},\\
C_{i} &=& \beta c_{i+1} c_{i+2} c_{i+4} + \gamma c_{i} c_{i+3} c_{i+4}.
\end{eqnarray}
For brevity, we have introduced the parameters
\begin{equation}
    \alpha = \sqrt{V_{012210}}, \ \beta = \sqrt{V_{023320}}, \ 
    \gamma = e^{i\theta} \sqrt{V_{014410}} \ ,
\end{equation}
given in terms of matrix elements (\ref{eq:matrix elements}) and $\theta = 2 \kappa^{2} g_{12}/g_{11}$. Amongst the terms omitted, \cref{eq:mr term 3,eq:mr term 5,eq:mr term 8} do not act directly on the root state, bringing only subleading contributions. While the term in \cref{eq:mr term 6} can directly act on the extreme root state, its contribution is suppressed in the vicinity of the thin-cylinder limit because of the prefactor being much smaller than the hopping term retained, Eq.~(\ref{eq:mr term 4}). 

Therefore, to a first order approximation, $H'_{\mathrm{MR}}$ is the correct effective Hamiltonian that captures the departure of the Pfaffian state from the root $|R_{0}\rangle$ in this geometry  (in Sec.~\ref{sec:dynamics} we will also investigate its dynamical properties to show that the model captures the properties of excited eigenstates). On the torus, our model preserves the $6$-fold ground state degeneracy of the MR Hamiltonian. Four of those, corresponding to root unit cell $1001$ and its translations, will become ``dressed", in analogy to the ground state of the cylinder Hamiltonian. The other two will remain inert, i.e. equal to $|101010\dots\rangle$ and $|010101\dots\rangle$ for any value of $L_{2}$ (since the hopping in \cref{eq:mr term 4} cannot produce new configurations).

To confirm the validity of the model in Eq.~(\ref{eq:cylinder model}) we performed several tests. First, we evaluated the overlap of the model's ground state with the full MR state, i.e., the ground state of the untruncated Hamiltonian. Fig.~\ref{FredkinOverlap} shows that this overlap is very high close to the thin-cylinder regime, with overlap on the order ${\sim}95\%$ at $L_{2} {\approx}7\ell_B$. As we are not exactly capturing the full state at a finite value of $L_2$, the overlaps naturally decay with system size (and vanish in the thermodynamic limit). Nevertheless, the fact that they remain very high and weakly dependent on system size in the range $L_2\lesssim 7\ell_B$ gives us confidence that the model captures the right physics, as will be further demonstrated below.

\begin{figure}[t] 
    \includegraphics[width=1\columnwidth]{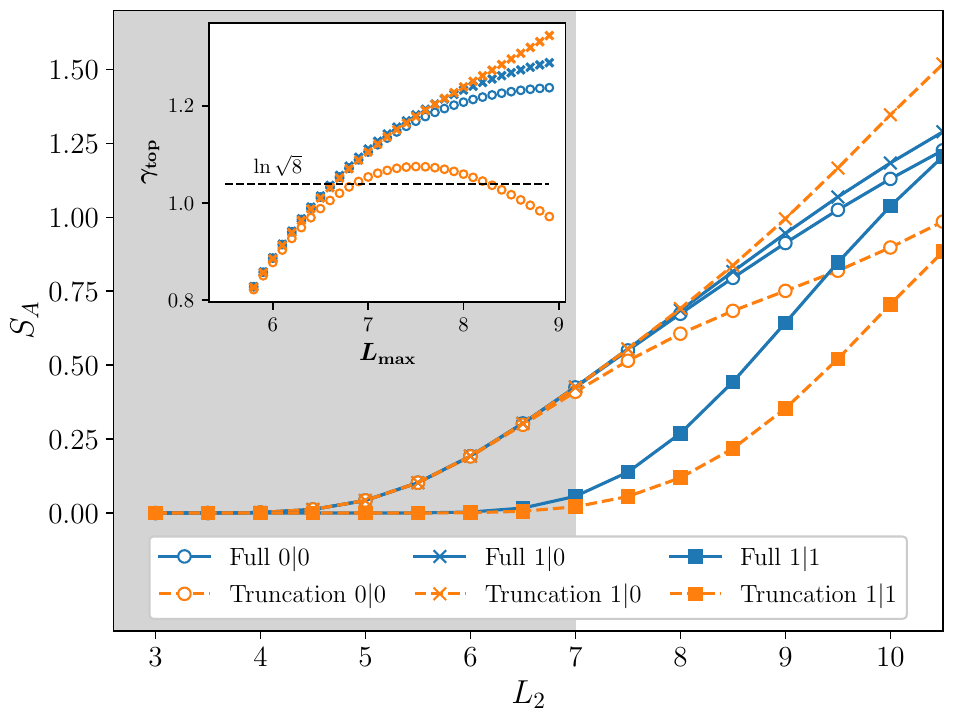}
	\caption{Bipartite entanglement entropy $S_{A}$ of the full MR state [i.e., the ground state of \cref{eq: MR parent hamiltonian}] and the ground state of the truncated model, \cref{eq:cylinder model}, as a function of cylinder circumference, $L_2$. Data is for $N_{e}{=}14$ electrons and $N_{\phi}{=}26$ magnetic orbitals. All types of bipartitions are considered: choosing subsystem $A$ to contain $N_\phi^{A}{=}11$ orbitals, the boundary looks like $0|0$; it can be seen that entanglement entropy here starts growing early on. The case $N_\phi^{A}{=}12$, corresponding to $1|0$, behaves similarly. By contrast,  choosing $N_\phi^{A}{=}13$ corresponds to the bipartition type $1|1$, where entropy grows much more slowly. The truncated model accurately captures the behavior of the model in the range of circumferences $L_{2}\lesssim 7\ell_{B}$ (shaded). The inset shows the topological entanglement entropy $\gamma_{\mathrm{top}}$, extracted numerically from $S_{A}$ using a linear fit over the interval $[5.5\ell_{B}, L_{\mathrm{max}}]$. Only the bipartitions $0|0$ and $1|0$ were used, as those scale correctly near the Tao-Thouless limit. Restricting ourselves to the range of validity for the truncated model, our $\gamma_{\mathrm{top}}$ estimate is within $20\%$ of the theoretical value of $\ln \sqrt{8}$.}
	\label{FredkinEntanglementEntropy}
\end{figure}
An example of a physical quantity that can be meaningfully scaled with system size and is sensitive to both local and nonlocal correlations is the bipartite entanglement entropy, $S_A$. We compute $S_A$ by choosing a bipartition in orbital space, i.e., the subsystem $A$ contains $N_\phi^A$ orbitals and the complementary subsystem $B$ contains the remaining $N_\phi{-}N_\phi^A$ orbitals~\cite{Li08}. Due to the Gaussian localization of the magnetic orbitals (\ref{eq:orbitals}), this roughly corresponds to the more conventional partitioning of the system in real space~\cite{Dubail12,Sterdyniak12}. The entanglement entropy is the von Neumann entropy, $S_A=-\mathrm{tr}\rho_A\ln\rho_A$, of the reduced density matrix $\rho_A=\mathrm{tr}_B|\psi\rangle\langle \psi|$ for the (truncated) MR ground state $|\psi\rangle$. In Fig.~\ref{FredkinEntanglementEntropy} we plot $S_A$ as a function of cylinder circumference, contrasting the full MR state with the ground state of the truncated model (\ref{eq:cylinder model}). The entanglement entropy of the MR state has been shown to scale linearly with the circumference of the cylinder~\cite{Zaletel12,Estienne15}, which is the ``area law" scaling expected in ground states of gapped systems~\cite{Eisert10}. We observe that this linear scaling is obeyed also by the truncated model for $L_2\lesssim 7\ell_B$. Furthermore, the subleading correction $\gamma_\mathrm{top}$ to the area law, $S_A = c L_2 - \gamma_\mathrm{top} + O(e^{-\xi/\ell_B})$, where $c$ is a constant and $\xi$ is the correlation length, is known to be a sensitive indicator of topological order as it arises due to fractionalized anyon excitations~\cite{Kitaev06a,Levin06}. As shown in the inset of Fig.~\ref{FredkinEntanglementEntropy}, in the range of validity of the truncated model we obtain $\gamma_\mathrm{top}$ within 20\% of the theoretically expected value $\ln \sqrt{8}$ for the MR case~\cite{Zaletel12}. 

Beyond the area-law regime, the entanglement entropy of the truncated model saturates, illustrating that the model is no longer able to capture the relevant correlations in the full MR state. Conversely, near the Tao-Thouless limit  $L_2{\lesssim} 4\ell_B$, there is practically no growth of entropy with $L_2$, as the ground state remains a product state. In the latter regime, $\gamma_{\mathrm{top}}{=}0$, illustrating that topological order is completely lost since the system is too narrow in the $y$ direction.

Finally, \cref{FredkinEntanglementEntropy} illustrates  the sensitivity of entanglement scaling with respect to the location of the bipartition. This is due to the cylinder ground state being dominated by the pattern $\dots 11001100 \dots$.  Given this form of the root state unit cell, there are three distinct types of locations where we could place the partition. If we partition between two occupied orbitals (i.e., $1|1$), at first order there will be no correlated hoppings across this boundary, leading to a very slow growth in entanglement, as indeed seen in \cref{FredkinEntanglementEntropy}. 
Instead, if we partition the system next to an empty orbital (i.e., at $0|0$ or $1|0$), then there will be correlated hopping across the boundary and the two subsystems can get entangled more easily. Note that this sensitivity to the location of the partition is not present in the Laughlin case~\cite{Nakamura2012}. Moreover, it is an artefact of being near the thin-cylinder limit where the ground state still possesses a crystalline-like density modulation, which becomes strongly suppressed in the isotropic 2D limit where the fluid is spatially uniform. Nevertheless, \cref{FredkinEntanglementEntropy} shows that our truncated model (\ref{eq:cylinder model}) successfully captures all the entanglement features of the full MR state in the regime $L_2\lesssim 7\ell_B$. In the next section, we show that the model (\ref{eq:cylinder model}) can be expressed as a well-known spin-1/2 chain model.

\section{Mapping to a deformed Fredkin chain}\label{sec:spinmodel}

Our effective Hamiltonian (\ref{eq:cylinder model}) is frustration-free and it has an exact zero-energy ground state, which is unique on a cylinder. To write down the ground state wave function and provide its intuitive representation, we map the model (\ref{eq:cylinder model}) to a deformed Fredkin chain~\cite{Salberger2016,Salberger2017, ChenXiao17}. The Fredkin model is a spin-$1/2$ analog of the Motzkin chain model~\cite{Bravyi12,Movassagh16,Movassagh17,ZhangZhao17}. As shown in Appendix~\ref{app:motzkin}, the Motzkin chain allows to generalize the construction from Ref.~\cite{Nakamura2012} and describe the $\nu{=}1/3$ Laughlin state over a larger range of cylinder circumferences.

\subsection{Spin mapping}

The mapping to spin-1/2 degrees of freedom is possible because we are only interested in the connected component of the ground state, i.e., the Krylov subspace spanned by states $(H_\mathrm{MR}^\prime)^n |11001100\dots 110011\rangle$, for an arbitrary integer $n$~\cite{MoudgalyaKrylov}. For our truncated model, the dimension of such a subspace does not exhaust the full Hilbert space dimension, allowing the mapping onto a spin-1/2 chain. This is an example of Hilbert space fragmentation~\cite{MoudgalyaReview} and it has previously been used to perform mappings of 2-body FQH Hamiltonians onto spin models~\cite{Moudgalya2019}. Thus, without any loss of generality, we can investigate both the ground state and dynamics by restricting to the Krylov subspace built from the Tao-Thouless  root pattern.

\begin{figure}[tb]
    \centering
    \includegraphics[width=\linewidth]{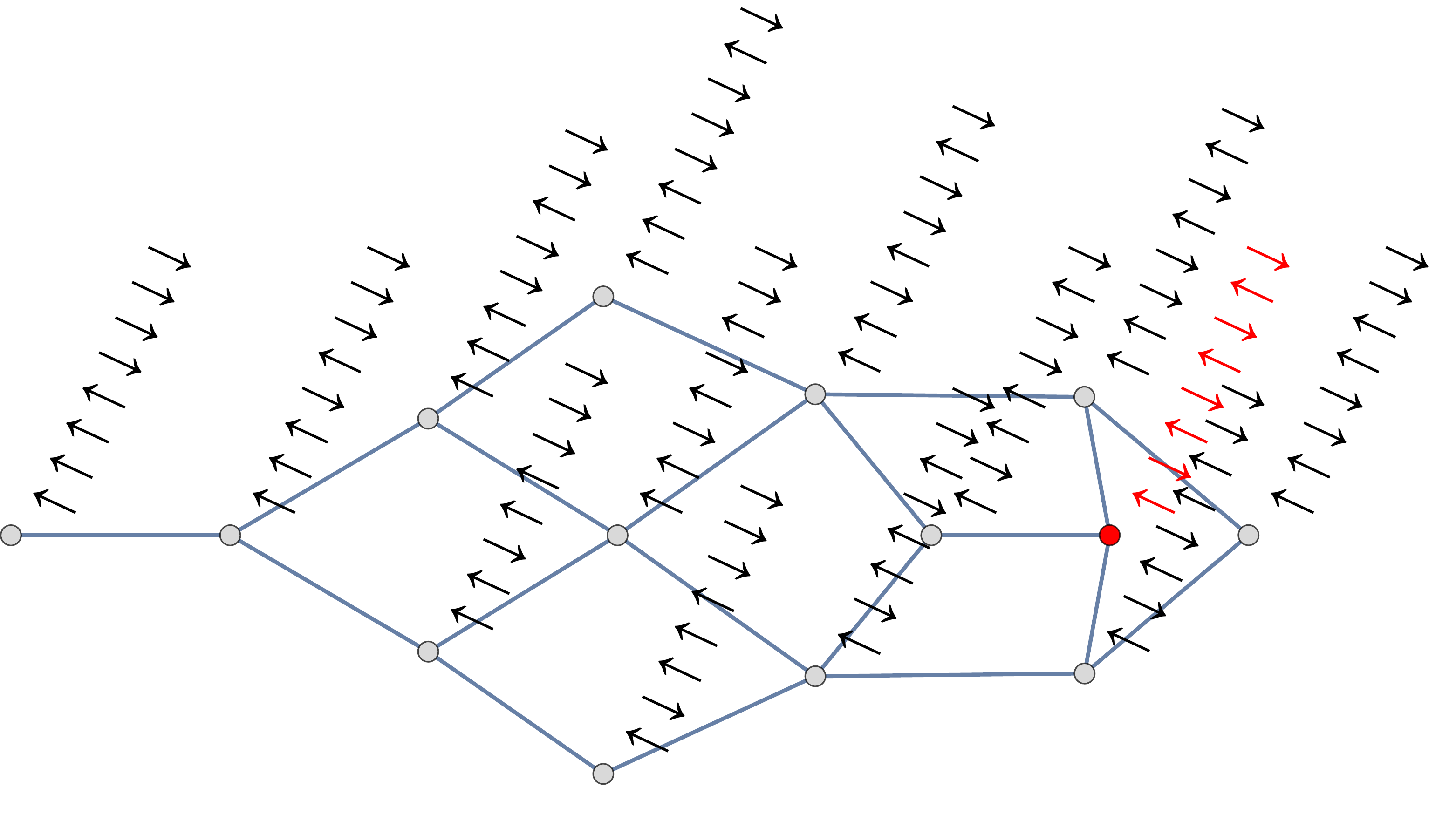}
    \includegraphics[width=\linewidth]{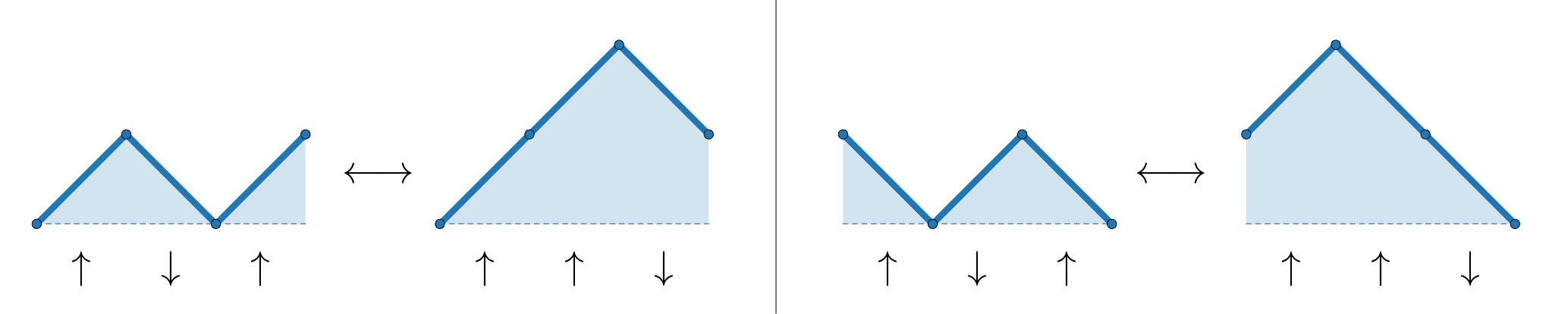}
    \caption{Adjacency graph of the truncated model $H_\mathrm{MR}^\prime$, Eq.~(\ref{eq:cylinder model}), in the connected sector containing the root state $(0)11001100110011(0)$ or, equivalently, $\uparrow\downarrow\uparrow\downarrow\uparrow\downarrow\uparrow\downarrow$ in the spin representation (marked in red). The graph vertices are product states that can be reached by repeated  application of the Hamiltonian. The edges denote the nonzero matrix elements of the Hamiltonian between respective vertices.  This example is for $N_e{=}8$ electrons, where the connected component of the root state contains only $14$ out of total $151$ states with the same momentum quantum number. Any basis configurations not present here are dynamically disconnected and cannot be reached from the ground state under the dynamics generated by $H_\mathrm{MR}^\prime$. The Fredkin moves (swaps with $\uparrow$ on the first or $\downarrow$ on the last site acting as a control qubit), as implemented by the Hamiltonian, are also shown in the bottom panels. Each move changes the area of the path by a constant value.}
    \label{fig: adjacency graph}
\end{figure}

To perform the mapping, start from the root state $110011 \dots 0011$ and pad it with one fictitious $0$ on each side.  This allows us to group sites in pairs of two, noticing that the only present pairs are $01$ and $10$. These are mapped to spins:
\begin{equation}
    |01\rangle \to \, \mid \uparrow \rangle \quad \mathrm{and} \quad |10\rangle \to \, \mid \downarrow \rangle.
\end{equation}
Thus, the root state maps to the antiferromagnetic (N\'eel) state $|(0)110011\dots0011(0)\rangle \to \, \mid \uparrow \downarrow \uparrow \downarrow \dots \uparrow \downarrow \rangle$. Since $H_\mathrm{MR}^\prime$ acts on a maximum of 5 consecutive sites at once, its equivalent acts on 3 consecutive spins. As discussed below, acting with $H_\mathrm{MR}^\prime$ on any product state that can be mapped to spins, i.e., consists of a sequence of 01 and 10 pairs, generates another product state that can be similarly mapped to spins. An example is presented in Fig.~\ref{fig: adjacency graph} which shows the connected sector of the adjacency graph of $H_\mathrm{MR}^\prime$ that contains the N\'eel state.
This shows that in this connected component of the Hilbert space, our model~(\ref{eq:cylinder model}) is equivalent to a spin-$1/2$ model. For simplicity, we will denote the number of spins by $N$, although it should be kept in mind this is equal to the number of electrons $N_e$.

The moves implemented by the $B$ and $C$ terms in Eq.~(\ref{eq:cylinder model}) are $\mid \uparrow \uparrow \downarrow \rangle \leftrightarrow \ \mid \uparrow\downarrow \uparrow \rangle$ and $\mid \uparrow \downarrow \downarrow \rangle \leftrightarrow \  \mid \downarrow \uparrow \downarrow \rangle$, i.e., they are controlled swaps of spins or Fredkin gates~\cite{NielsenChuang}, as illustrated in \cref{fig: adjacency graph}. Note that the $A$ term in our subspace is redundant, as this subspace does not contain any $\dots 111 \dots$ patterns.  The resulting spin Hamiltonian is a sum of local projectors and can be written as 
\begin{align}\label{eq:fredkinmodel}
    H_{F} &= \sum_{i=0}^{N-3} \mathcal{P}^{\uparrow}_{i}P^{\varphi(\tau)}_{i+1,i+2} + \mathcal{P}^{\varphi(\tau)}_{i,i+1}\mathcal{P}^{\downarrow}_{i+2} 
\end{align}
where the single-spin projector $\mathcal{P}_i^\sigma{=}|\sigma_i\rangle\langle\sigma_i|$ projects onto a local spin pointing in the direction $\sigma=\uparrow,\downarrow$. The two-spin projector 
$\mathcal{P}^{\varphi(\tau)}_{i,i+1} = |\varphi(\tau)_{i,i+1}\rangle\langle\varphi(\tau)_{i,i+1}|$ projects onto the deformed superposition state  
\begin{eqnarray}\label{eq:deformation parameter}
 |\varphi(\tau)_{i,i+1}\rangle &=& \mid\uparrow_{i}\downarrow_{i+1}\rangle -\tau\mid\downarrow_{i}\uparrow_{i+1}\rangle, \\
    \tau = -\gamma / \beta &=& -2\exp \left( -2\kappa^{2}\frac{1-ig_{12}}{g_{11}} \right).
\end{eqnarray}
The model (\ref{eq:fredkinmodel}) is our central result of this section. It can be recognized that this model corresponds to a (colorless) deformed Fredkin chain from Ref.~\cite{Salberger2017}. 
Note that the boundary Hamiltonian terms from Ref.~\cite{Salberger2017}, i.e.,  $ H_{\partial} = \mathcal{P}^{\uparrow}_{0} + \mathcal{P}^{\downarrow}_{N-1} $, have been omitted because our subspace has the first $\uparrow$ spin and the last $\downarrow$ spin frozen.

For convenience, we note that $H_F$ can be equivalently expressed in terms of usual Pauli spin operators:
\begin{equation}\label{eq:H_Pauli}
    H_{\mathrm{F}} = \sum_{j=0}^{N_{e}-3} \biggl(  \bigl(\frac{1}{2}\mathds{1} + S^{z}_{j} \bigr) h_{j+1,j+2}(\tau) + h_{j,j+1}(\tau) \bigl(\frac{1}{2}\mathds{1} - S^{z}_{j+2} \bigr) \biggr),
\end{equation}
where 
\begin{align}
h_{j,j+1}(\tau) &= \frac{1+|\tau|^{2}}{4} (\mathds{1} - 4 S^{z}_{j}S^{z}_{j+1}) + \frac{1-|\tau|^{2}}{2} (S^{z}_{j}-S^{z}_{j+1})\nonumber \\ &  - 2 \mathrm{Re}(\tau)(S^{x}_{j}S^{x}_{j+1}+S^{y}_{j}S^{y}_{j+1}) \nonumber \\ & - 2\mathrm{Im}(\tau)(S^{x}_{j}S^{y}_{j+1}-S^{y}_{j}S^{x}_{j+1}).
\end{align}

Note that outside the connected component of the ground state, the mapping can be extended by defining the additional composite degrees of freedom: $|11\rangle \to |{+}\rangle$, $|00\rangle \to |{-}\rangle$. In this mapping, the constrained dynamics of the model resembles that of fractonic models in Refs.~\cite{Nandkishore19,Pai19,Pai20,Sous20}, which can lead to different thermalization properties in different Krylov fragments~\cite{MoudgalyaKrylov}.

\subsection{The ground state}

The ground state of the Fredkin chain is a weighted superposition of ``Dyck paths"  of length $N$ (the set of which we denote $\mathcal{D}_{N}$). These are product state configurations with $S^{z}_{\mathrm{total}}=0$ and $\sum_{i=0}^{k}S^{z}_{i} \geq 0$ for all $k$. The last condition is equivalent to the number of spin $\uparrow$ sites always being greater than or equal to the number of spin $\downarrow$ sites, as we go through the chain from left to right. The paths in $\mathcal{D}_{N}$ can be interpreted graphically, as a ``mountain range" where each $\uparrow$ corresponds to an upward slope, while $\downarrow$ corresponds to a downward slope. The Dyck constraint is equivalent to the height of these graphs starting and ending at zero, and always staying positive. The weight of each path (configuration) $p$ in the ground state $|\psi_{0}\rangle$ will be determined by the area $A(p)$ under the mountain range:
\begin{equation} \label{eq:fredkin gs}
    |\psi_{0}\rangle = \mathcal{N}^{-1} \sum_{p \in \mathcal{D}_{N}} \tau^{A(p)/2} \, |p\rangle.
\end{equation}
According to the phase diagram of the Fredkin spin model, for $|\tau|<1$ the entanglement entropy of the ground state is bounded (obeys area law), whereas $|\tau|=1$ is a critical point where the scaling becomes logarithmic in system size $\sim \log N$.

Ref.~\cite{Salberger2016} also discusses the subtleties of the spin model with periodic boundary conditions. For $|\tau|=1$, the ground state degeneracy scales linearly with $N$, with zero modes in every $S^{z}_{\mathrm{total}}$ sector. However, upon decreasing the deformation away from the critical point we find that the extensive degeneracy disappears -- only 4 zero modes survive. Two of them are in the sectors $S^{z}_{\mathrm{total}} = \pm N/2$, corresponding to the inert states $|{\uparrow}\rangle ^{\otimes N}$ and $|{\downarrow}\rangle ^{\otimes N}$ (or the Fock states $|101010\dots\rangle$ and $|010101\dots\rangle$). The other two are in the sector $S^{z}_{\mathrm{total}} = 0$ and will correspond to the root unit cells $1001$ and $0110$; the two remaining translations break our spin mapping but can be obtained by shifting every orbital by one position and then applying the mapping. All other zero modes disappear because deformed Fredkin ground states are constructed using the Dyck path area as in \cref{eq:fredkin gs}, which is only well defined when $S^{z}_{\mathrm{total}}=0$. These results are in agreement with the $6$-fold degeneracy found in the fermionic model.

With this understanding, we can map back to the fermionic ground state. All Dyck paths can be obtained from the root configuration $\mid \uparrow \downarrow \uparrow \dots \downarrow \uparrow \downarrow \rangle$ by exchanging a number of $\downarrow$ with an equal number of $\uparrow$ further along the chain. In the fermionic picture, this is equivalent to performing a number of ``squeezes'', i.e., applications of the operator 
\begin{eqnarray}\label{eq:squeezing}
 \hat{S}_{k,d} = c^{\dagger}_{2k}c^{\dagger}_{2(k+d)-1}c_{2(k+d)}c_{2k-1},
\end{eqnarray}
with $d,k > 0$. Similar structure exists in the Laughlin state~\cite{Rahmani20, Kirmani2021}. The resulting states are in one-to-one correspondence with those in $\mathcal{D}_{N}$, and we will denote their set by $\mathcal{D}_{N}'$. Every configuration $s$ in $\mathcal{D}'_{N}$ can be obtained from a number of $n(s)$ squeezes applied to the root.
\begin{equation}
    s \in \mathcal{D}_{N}'  \Longleftrightarrow |s\rangle = \prod_{i=1}^{n(s)} \hat{S}_{k_i,d_i} \, |110011\dots0011\rangle
\end{equation}
The weight of such a basis state in the ground state is now determined by the total distance squeezed $D(s) = \sum_{i=1}^{n(s)} d_i$, which is equivalent to the previous definition (\ref{eq:fredkin gs}) expressed in terms of the area under the Dyck path.
\begin{equation} \label{eq:fredkin gs fermions}
    |\psi_{0}\rangle = \mathcal{N}^{-1} \sum_{s \in \mathcal{D}_{N}'} \tau^{D(s)/2} \, |s\rangle.
\end{equation}

\subsection{Matrix-product state representation} \label{app:mps fredkin}

Along with the undeformed chain ($\tau{=}1$), Ref.~\cite{Salberger2016} introduced a matrix-product state (MPS) representation for its ground state. The associated MPS matrices have bond dimension $\chi=N/2+1$, where $N$ is the number of spins:
\begin{equation} \label{eq:MPS critical}
    M^{\uparrow}_{jk} = \delta_{j+1,k} \quad \mathrm{and} \quad  M^{\downarrow} = (M^{\uparrow})^{T}.
\end{equation}
As we are working with open boundary conditions, we use the boundary vectors $v_{L}=v_{R}^{T}$ with $(v_{L})_{j}=\delta_{j,0}$. This MPS can be directly extended to the deformed chain, where we need to introduce the deformation parameter in the following way:
\begin{equation} \label{eq:MPS deformed}
    (M^{\uparrow}_{\tau
    })_{jk} = \tau^{j/2}\, \delta_{j+1,k} \quad \mathrm{and} \quad  M^{\downarrow}_{\tau} = (M^{\uparrow}_{\tau})^{T}
\end{equation}
This holds for any $\tau \in \mathbb{C}$ so it can be used for anisotropic states as well. Therefore the Fredkin ground state can be written as:
\begin{equation}
    |\psi_{0}\rangle = \mathcal{N}^{-1/2}\, v^{T} M_{\tau,0}M_{\tau,1} \dots M_{\tau,N-1} \,v, 
\end{equation}
where the MPS tensor is given by
\begin{equation}
    M_{\tau,j} = \begin{pmatrix}
    0 & |{\uparrow_{j}}\rangle & 0 & 0 & \dots \\
    |{\downarrow_{j}}\rangle & 0 & \tau^{1/2}|{\uparrow_{j}} \rangle & 0 & \\
    0 & \tau^{1/2}|{\downarrow_{j}}\rangle & 0 & \tau|{\uparrow_{j}}\rangle & \\
    0 & 0 & \tau|{\downarrow_{j}}\rangle & 0 & \\
    \vdots &  &  &  & \ddots 
    \end{pmatrix}.
\end{equation}
For e.g. $N{=}6$ this expression gives:
\begin{align}
    |\psi_{0}^{N=6}\rangle &= |{\uparrow\downarrow\uparrow\downarrow\uparrow\downarrow}\rangle + \tau\big(|{\uparrow\downarrow\uparrow\uparrow\downarrow\downarrow}\rangle+|{\uparrow\uparrow\downarrow\downarrow\uparrow\downarrow}\rangle\big) \nonumber \\ &+ \tau^{2}|{\uparrow\uparrow\downarrow\uparrow\downarrow\downarrow}\rangle + \tau^{3}|{\uparrow\uparrow\uparrow\downarrow\downarrow\downarrow}\rangle
\end{align}
which indeed agrees with Eq.~(\ref{eq:fredkin gs fermions}). We note that alternative  tensor network representations of the Fredkin ground states have been discussed in the literature \cite{Alexander21}. 

\begin{figure}[tb]
    \centering
    \includegraphics[width=\linewidth]{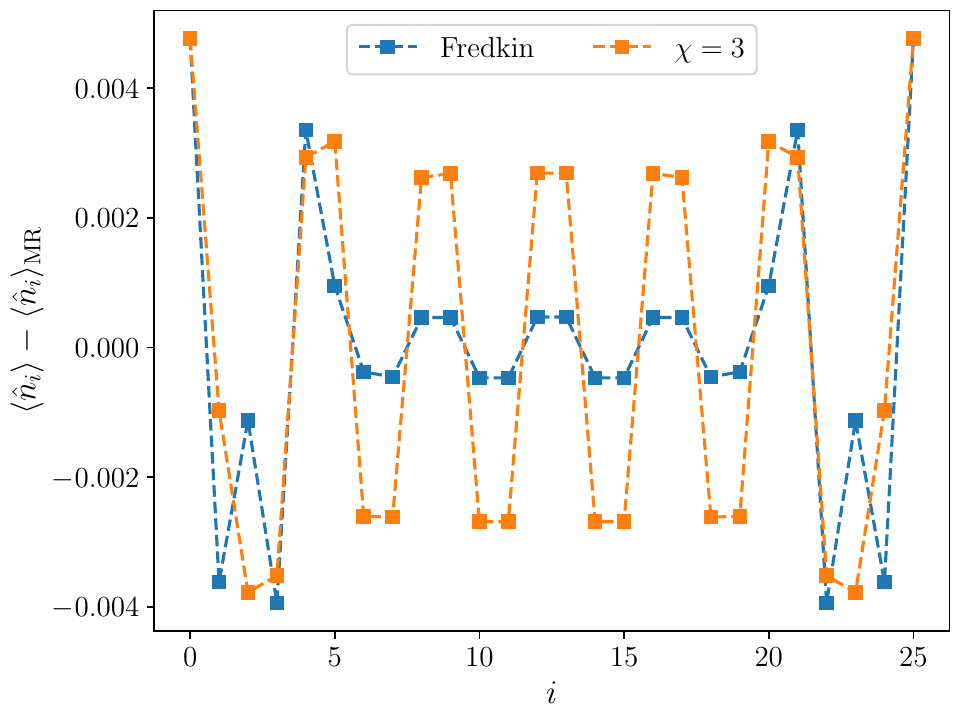}
    \caption{Comparison between the average orbital density of the MR state and different truncated states: the Fredkin state and the state obtained by truncating the Fredkin MPS at $\chi{=}3$. The system has $N_e{=}14$ electrons and the cylinder circumference is $L_{2} {=} 7\ell_{B}$. The truncated states deviate slightly from the charge density wave pattern of the MR state.}
    \label{fig: orbital density comparison}
\end{figure}

Furthermore, the MPS representation above is able to capture the critical point at $|\tau|=1$, which is precisely the reason behind $\chi$ increasing linearly with system size. This limits our ability to extract thermodynamic limit behaviour in this phase using the MPS tensors from \cref{eq:MPS critical,eq:MPS deformed}. However, the regime of interest for this paper is $|\tau| \lesssim 0.4$ (i.e., $L_{2} \lesssim 7\, \ell_{B}$), which is far from the critical point. Hence, it is possible to describe the ground state with high accuracy by truncating to a finite bond dimension. Consider the following tensors:
\begin{equation}\label{eq:MPSapprox}
    M_{\tau,j}^{(\chi=3)} = \begin{pmatrix}
    0 & |{\uparrow_{j}}\rangle & 0 \\
    |{\downarrow_{j}}\rangle & 0 & \tau^{1/2}|{\uparrow_{j}} \rangle \\
    0 & \tau^{1/2}|{\downarrow_{j}}\rangle & 0
    \end{pmatrix}.
\end{equation}
For a chain with even number of spins, this MPS yields the following simple ground state:
\begin{equation}\label{eq: circuit gs}
    |\psi_{0}^{(\chi=3)}\rangle = |{\uparrow}\rangle \big( |{\downarrow \uparrow}\rangle + \tau |{\uparrow \downarrow}\rangle \big)^{\otimes \frac{N-2}{2}} |{\downarrow}\rangle.
\end{equation}
With a fixed $\chi$, it is straightforward to analytically calculate the  behavior of relevant quantities in the thermodynamic limit by using the MPS transfer matrix. The average orbital density takes the form:
\begin{equation}\label{eq:orbital_density}
    \langle \hat{n}_{4j/4j+1} \rangle = \frac{1}{1+\tau^{2}} \, ,
    \quad \langle \hat{n}_{4j+2/4j+3} \rangle = \frac{\tau^{2}}{1+\tau^{2}}.
\end{equation}
As expected, this resembles a CDW pattern, which in this approximation (and also in the full Fredkin ground state state) is predicted to disappear at $|\tau|=1$, corresponding to $L_{2} {\approx} 10.7\ell_{B}$ (outside the range of validity of the truncated model). Figure~\ref{fig: orbital density comparison} shows a comparison of orbital density between the MR state, the Fredkin state and the $\chi{=}3$ approximation above. At $L_{2} {=} 7\ell_{B}$ the two truncated states still capture the CDW pattern, with the Fredkin state showing more accurate results. Since this approximate state can be written in the tensor product form above, the density-density correlations decay to zero with a finite correlation length.

\subsection{Quantum algorithm for preparing the ground state and its implementation on IBM quantum processor}\label{sec:IBM}

The simple structure of the MPS wave function in the above approximation (for $\chi{=}3$) is amenable to implementation on noisy intermediate-scale quantum devices. Indeed, all states in the superposition can be obtained from a direct-product root pattern by only one layer of one- and two-qubit gates. Furthermore, the parameters of the circuit can be determined analytically without the need for any classical or hybrid optimization, which allows for direct implementation on a large number of qubits. 

\begin{figure}[b]
    \centering
    \includegraphics[width=0.6\linewidth]{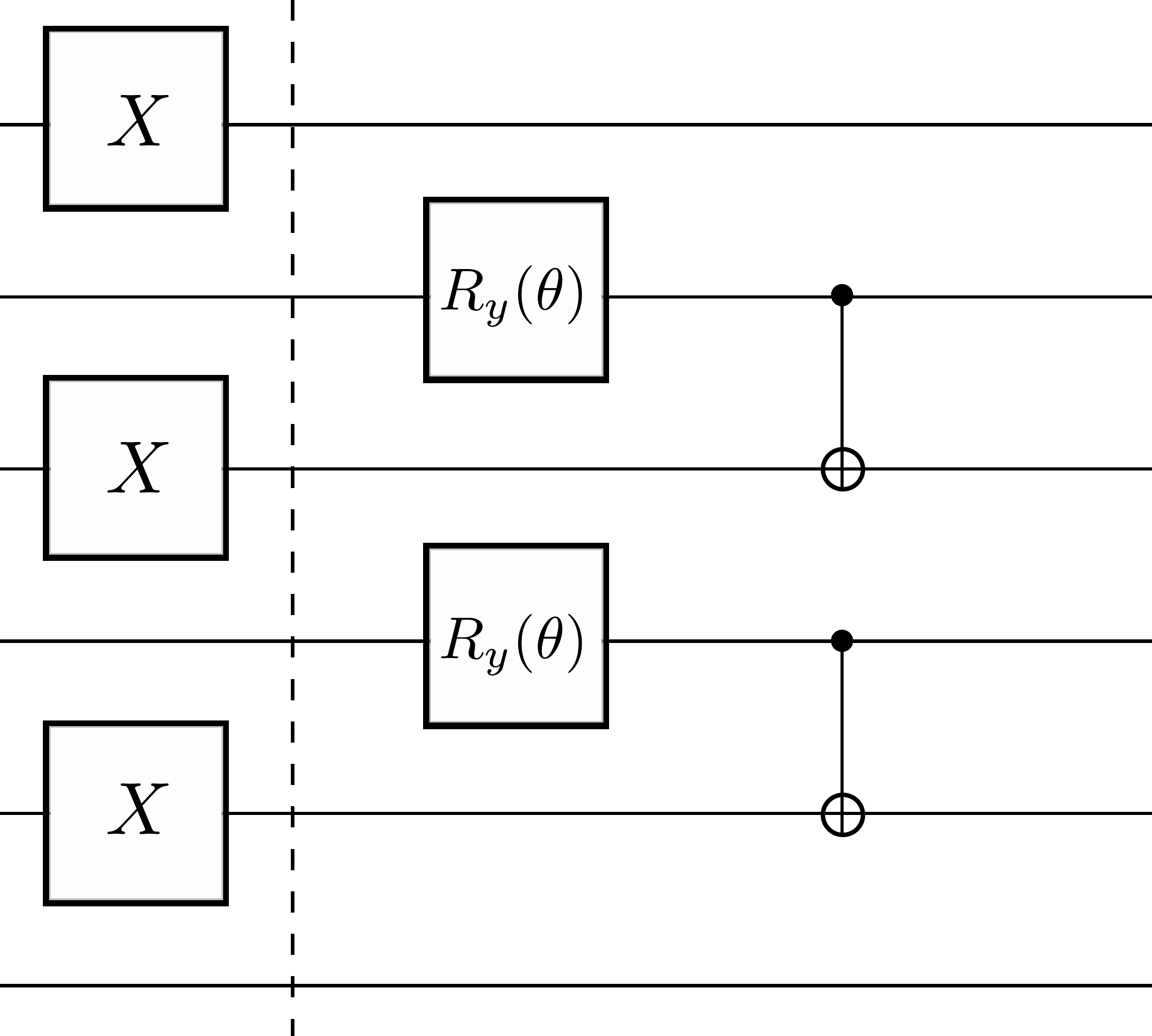}
    \caption{The structure of the quantum circuit for six qubits. The $X$ gates create the root patterns, and the rotations and {\footnotesize CNOT}s implement the action of MPS matrices in Eq.~(\ref{eq:MPSapprox}). }
    \label{fig: circuit}
\end{figure}

The structure of the quantum circuit is shown in Fig.~\ref{fig: circuit}. If we choose the angle $\theta$ to be equal to
\begin{equation}
\theta=2\arctan(\tau),
\end{equation}
the $y$-rotation creates a superposition $|{\uparrow}\rangle +\tau |{\downarrow}\rangle$ and the {\footnotesize CNOT} then changes the state of the two qubits to $|{\downarrow\uparrow}\rangle +\tau |{\uparrow\downarrow}\rangle$.  
As a quick check, we executed the circuit on the ibmq\_mumbai device, a 27-qubit processor with quantum volume 128. This implementation was carried out using the Qiskit package. In this similuation, we used $N=26$, with the initial and final qubits held in trivial up and down states. Notably, we refrained from employing any error mitigation techniques, and we deliberately incorporated qubits and couplings from the device with lower quality. Our simulation utilized a mere couple of thousand shots to ascertain bitstring probabilities in the computational basis. We found very good agreement of the measured orbital densities with the analytical results of Eq.~\eqref{eq:orbital_density}, save for a few instances where gate calibrations during the simulation were imperfect.

\begin{figure}[tb]
    \centering
    \includegraphics[width=\linewidth]{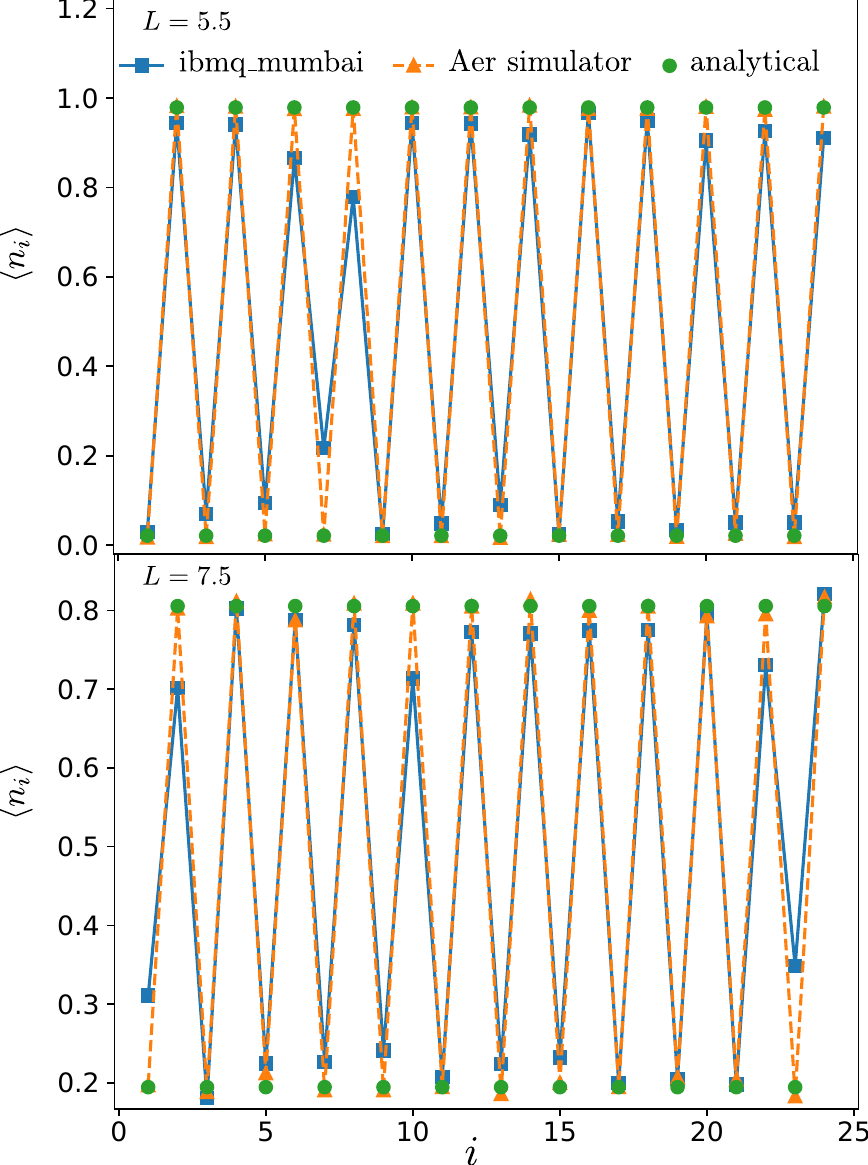}
    \caption{The orbital density from the quantum circuit. There is good agreement between the results of Eq.~\eqref{eq:orbital_density}, classical simulation of the circuit using IBM Aer simulator and quantum implementation on the ibmq\_mumbai device. The IBM data was obtained on 31/8/2023 at 2:02 PM. }
    \label{fig: orbital density}
\end{figure}

The results for the orbital density are shown in Fig.~\ref{fig: orbital density}. We used 2048 shots in both the quantum execution and the simulation of the circuit with IBM's Aer simulator. The Aer simulation is in excellent agreement with \cref{eq:orbital_density}, with slight differences due to the finite number of shots. There is also good agreement with the quantum device, except for a few qubits. 
Despite its simplicity, the approximate ground state prepared above can serve as a valuable starting point for exploring the dynamics of the MR phase on quantum computers.

\section{Quench dynamics} \label{sec:dynamics}

Given that our Fredkin model (\ref{eq:fredkinmodel}) was constructed by focusing on the root state and it does not represent a truncation of the MR Hamiltonian according to a decreasing order of magnitude, it is not obvious that the excited spectrum necessarily matches that of the full MR Hamiltonian. To demonstrate the correspondence of key physical properties between the two spectra, in this section we focus on the dynamical response of the Fredkin model. In particular, we study geometric quench~\cite{Liu18} to probe the compatibility between the two models, which was previously used to a similar effect in the  $\nu{=}1/3$ Laughlin case~\cite{Kirmani2021}. 

Geometric quench is designed to elicit the dynamical response of the Girvin-MacDonald-Platzman (GMP) collective mode~\cite{Girvin85,Girvin86}, which is present in all known gapped FQH states, including the MR state~\cite{Bonderson11NF, Moller11,Yang12b,Papic12, Repellin2014}. In the long-wavelength limit $k\ell_B {\to} 0$, the GMP mode forms a quadrupole degree of freedom that carries angular momentum $L{=}2$ and can be represented by a quantum metric~\cite{Haldane11}. In this respect,  the $k\ell_B {\to} 0$ limit of the GMP mode has formal similarity with the fluctuating space-time metric in a theory of quantum gravity~\cite{bergshoeff2013zwei,bergshoeff2018gravity}
and it is sometimes referred to as ``FQH graviton"~\cite{Yang12b,Golkar2016}.  It was shown that the quantum metric fluctuations can be exposed by introducing anisotropy which breaks rotational symmetry of the system~\cite{Liu18, Lapa19}.  Such geometric quenches induce coherent dynamics of the FQH graviton, even though the latter resides in the continuum of the energy spectrum, making it a useful probe of physics beyond the ground state considered thus far.  

\subsection{Spectral function}

The GMP mode, to a high accuracy, can be generated by a simple ansatz called the ``single-mode approximation''~\cite{Girvin85,Girvin86}:
the state belonging to the mode with momentum $\mathbf{k}$ is obtained by acting with the projected density operator, $\bar\rho_\mathbf{k}$, on the ground state, i.e., $|\psi_\mathbf{k}^\mathrm{GMP}\rangle = \bar \rho_\mathbf{k}|\psi_0\rangle$. Thus, the GMP states are automatically orthogonal to the ground state as they live in different momentum sectors. However, in practice, it is more convenient to study dynamics within the $\mathbf{k}=0$ sector of the ground state. This is the case with the geometric quench setup, described in Sec.~\ref{sec:geometricquench} below. Thus, in order to identify the relevant GMP state in the  $\mathbf{k}=0$ sector, possibly hidden in the continuum of the energy spectrum, we need a different tool. We identify the long-wavelength limit of the GMP mode using the following spectral function~\cite{Liu18,Liou19}:
\begin{align} \label{eq:spectral fn}
    I(\omega) &= \sum_{n} |\langle \psi_{n} | \hat{O} | \psi_{0} \rangle|^{2} \delta(\omega-\omega_{n}),
\end{align}    
where $\hat{O}$ is a 3-body operator with quadrupolar $x^2-y^2$ symmetry, given in Ref.~\cite{Liou19}, and the sum runs over (in principle, all) energy eigenstates $|\psi_n\rangle$ with energies $\omega_n$, measured relative to the ground state energy $\omega_0$.  In second quantization, the matrix element of $\hat O$ is 
\begin{align} \label{eq:spectral fn2}
    O_{j_{1}j_{2}j_{3}j_{4}j_{5}j_{6}} &=  \delta_{j_{1}+j_{2}+j_{3},j_{4}+j_{5}+j_{6}} \left(\sum j_{i}^{2} - \frac{1}{6}\big(\sum j_{i}\big)^{2}\right) \nonumber \\&\times (j_{1}-j_{2})(j_{2}-j_{3})(j_{1}-j_{3}) \nonumber \\
    &\times (j_{4}-j_{5})(j_{5}-j_{6})(j_{4}-j_{6}) \nonumber \\ &\times \exp \left[ -\frac{\kappa^{2}}{2}\left(\sum j^{2}_{i} - \frac{1}{6}(\sum j_{i})^{2}\right) \right],
\end{align}
which allows to readily evaluate Eq.~(\ref{eq:spectral fn}). Note that in this section we consider the spectral function for an \emph{isotropic} system, hence there is no metric dependence in Eq.~(\ref{eq:spectral fn2}). As before, the matrix element given here is derived for cylinder geometry and appropriate modifications are needed to make it compatible with torus boundary conditions, as explained in Sec.~\ref{sec:secondquantized}.

\begin{figure}[tb] 
    \includegraphics[width=1\columnwidth]{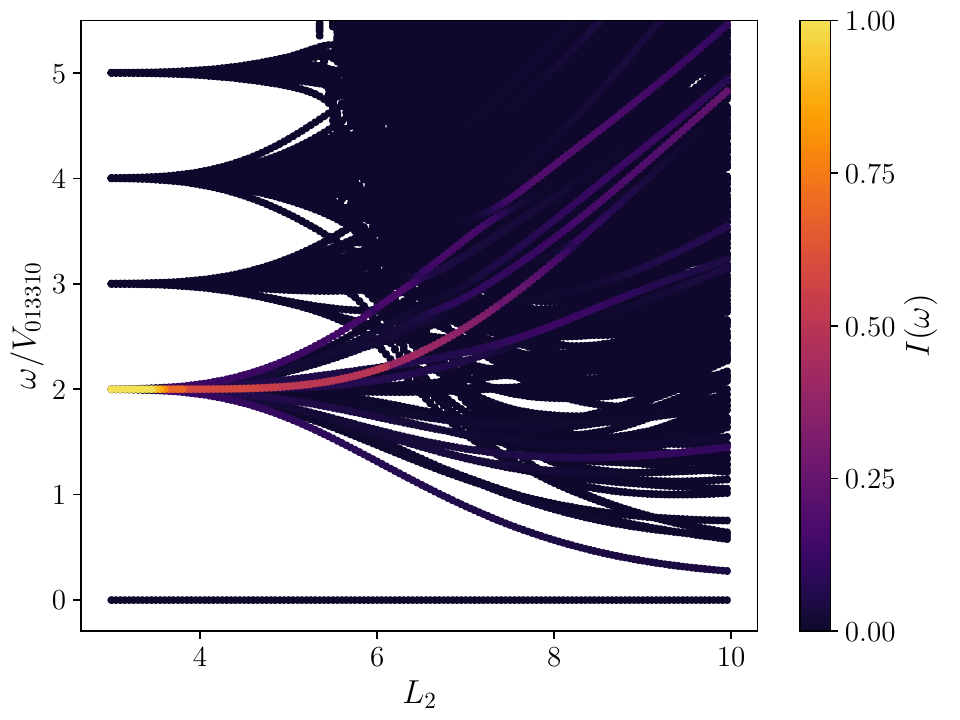}
    \includegraphics[width=1\columnwidth]{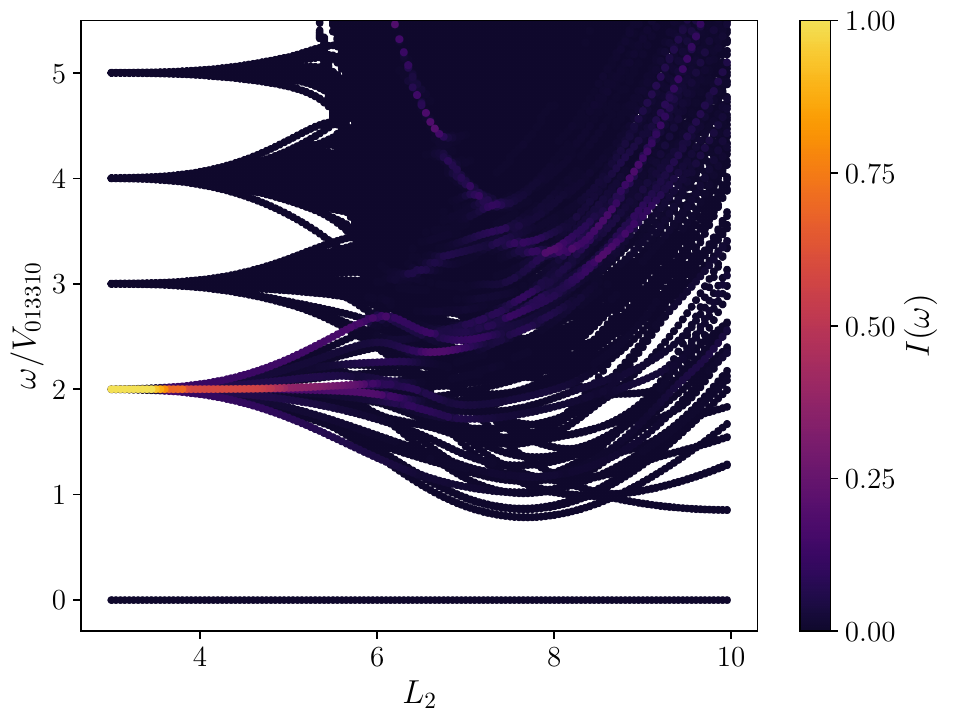}
	\caption{Evolution of the spectral function $I(\omega)$ in \cref{eq:spectral fn} in the Fredkin model (top) and the full MR model (bottom), as a function of cylinder circumference $L_2$. 
    The peak(s) in the spectral function are identified with the long-wavelength limit of the GMP mode, i.e., the FQH graviton. 
    System size is $N_{e}{=}10$ electrons, $N_{\phi}{=}18$ flux quanta. }
	\label{fig:spectral fn fredkin}
\end{figure}

In \cref{fig:spectral fn fredkin} we plot the evolution of the spectral function $I(\omega)$ as the cylinder circumference is varied from the Tao-Thouless limit towards the isotropic 2D limit, in both the untruncated and Fredkin models. We see there is good agreement between the two models for $L_2{\lesssim}7\ell_B$, i.e., across the same range where we previously established high overlap between the ground states of the two models. For larger circumferences, it becomes impossible to adiabatically track the evolution of the graviton peak in $I(\omega)$ due to multiple avoided crossings in \cref{fig:spectral fn fredkin}. The graviton resides in the continuum of the spectrum and it is not protected by a symmetry of the Hamiltonian, hence its support over energy eigenstates may undergo complicated  ``redistribution" as the geometry of the system is varied. In particular, away from the Tao-Thouless limit, there is also a clear splitting of spectral weight between several energy eigenstates, suggesting that the graviton degree of freedom may not correspond to a single eigenstate in this regime.  

\subsection{Geometric quench}\label{sec:geometricquench}

Given the complex evolution of the spectral function in \cref{fig:spectral fn fredkin} when interpolating between the isotropic 2D limit and the thin-cylinder limit we are interested in, it is natural to inquire if the graviton oscillations observed in the Laughlin case in Refs.~\cite{Liu18,Kirmani2021} persist in the MR  case and what their origin may be. In this section we analyze the geometric quench dynamics in the thin-cylinder limit and establish that it corresponds to a linearized bimetric theory of Gromov and Son~\cite{Gromov17}. This shows that, despite the simplicity of our model (\ref{eq:fredkinmodel}), it is successful at capturing a nontrivial many-body effect of a 2D FQH system away from equilibrium.

The geometric quench setup assumes that electrons are described by an arbitrary mass tensor $g_{ab}$, with $a,b=1,2$. The mass tensor must be symmetric and unimodular ($\mathrm{det}\; g = 1$)~\cite{Haldane11}, hence we can generally write it as $g=\exp(\hat Q)$ where $\hat Q= Q (2\hat{d}_a \hat{d}_b - \delta_{a,b})$ is a Landau-de Gennes order parameter and $\hat{\mathbf{d}}=(\cos(\phi/2), \sin(\phi/2))$ is a unit vector~\cite{Maciejko2013}. Parameters $Q$ and $\phi$ intuitively represent the stretch and rotation of the metric, respectively, with $Q{=}\phi{=}0$ corresponding to the isotropic case. Under Landau-level projection, the interaction matrix elements acquire explicit dependence on $g$, as can be seen in Eq.~(\ref{eq:matrix elements}). For $g$ close to the identity (i.e., at weak anisotropy), the topological gap is robust and the MR state remains a zero-energy ground state. We assume the initial state before the quench to be the isotropic MR state with $g=\mathds{1}$. At time $t=0$, the anisotropy in the Hamiltonian is instantaneously changed and, for simplicity, we assume the new metric to be diagonal, $g = \mathrm{diag}[g_{11},g_{22}]$, with $g_{11}\neq g_{22}$. The deformed $g_{11}$ (and, therefore, $g_{22}$) should be sufficiently close to unity such that the equilibrium system is still in the MR phase.    

\begin{figure}[tb]
\includegraphics[width=0.96\columnwidth]{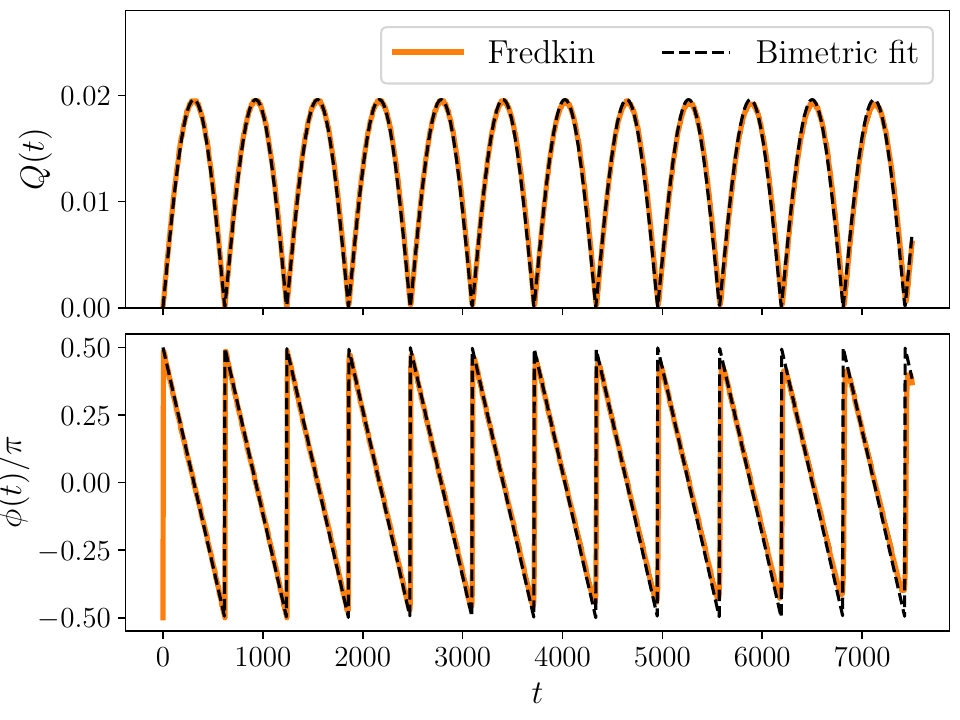}
	\caption{Geometric quench dynamics in the Fredkin model (\ref{eq:fredkinmodel}). The system size is $N_{e}{=}8$, $N_{\phi}{=}14$, and the circumference is $L_2 = 3.6\ell_{B}$. The system is initialised in the isotropic ground state and then time-evolved by the anisotropic Hamiltonian with $Q{=}0.01$. The resulting dynamics is in excellent agreement with the linearized bimetric theory, shown by dashed lines. The slight disagreement between the two at late times slow decay comes from the spectral weight in Fig.~\ref{fig:spectral fn fredkin} spread over more than a single energy eigenstate. }
    \label{fig:fredkin geometric quench}
\end{figure}

From \cref{eq:fredkin gs fermions} we can directly extract the first order corrections to the root state, $|R_0\rangle\equiv |11001100\ldots 0011\rangle$. These are given by states where only one squeezing, Eq.~(\ref{eq:squeezing}), is applied at a minimal distance:
\begin{align}
    |\psi_{0}\rangle \approx \big(1-\tau \sum_{i} \hat{S}_{i,1}\big) |R_{0}\rangle.
\end{align}
Substituting the deformation parameter in Eq.~(\ref{eq:deformation parameter}) and assuming $\exp(-2\kappa^{2})$ and the metric anisotropy $Q,\phi$ to be small, we get
\begin{align}\label{eq:psi0}
    |\psi_{0}\rangle \approx  |R_{0}\rangle - 2\exp\big[ -2\kappa^{2}\big(1-Qe^{i\phi} \big) \big] \sum_{i} \hat{S}_{i,1}|R_{0}\rangle,
\end{align}
where we used $g_{11}=\cosh Q$, $g_{12}=\sinh Q e^{i\phi}$ and therefore $(1-ig_{12})/g_{11}\approx 1-i Q e^{i\phi}$. 
On the other hand, the graviton state is approximated by:
\begin{align}
    |\psi_{g}\rangle &= \hat{O} |\psi_{0}\rangle \propto e^{-14\kappa^{2}/3}\bigg(\sum_{i} \hat{S}_{i,1}|R_{0}\rangle + O\big(e^{-2\kappa^{2}}\big)\bigg). 
\end{align}
Note that $|\psi_{g}\rangle$ is orthogonal to $|\psi_{\mathrm{0}}\rangle$. From here, we deduce the graviton root state, 
\begin{align} \label{eq:graviton root state}
    |R_{g}\rangle &= \sum_{i} \hat{S}_{i,1}|R_{0}\rangle \nonumber \\
    &= |1011010011\dots\rangle + |1100101101\dots\rangle + \dots,
\end{align}
which is proportional to the first order squeezes. This is identical to the MR ground state first-order correction to the root state and, in some sense, it is the simplest translationally invariant quadrupole structure that we can impose on top of it, creating quadrupoles of the form $-++-$ in each unit cell.

From the graviton root state, we can deduce the geometric quench dynamics up to first order in $\exp(-2\kappa^{2})$. Assuming, for simplicity, that the post-quench Hamiltonian has the metric $g_{11}=\exp(A)\approx 1+A$ and $g_{12}=0$, the initial state is given by
\begin{align}\label{eq:timeevolvedpsi}
    |\psi(t=0)\rangle = |\psi_{0}^{\mathrm{iso}}\rangle \approx |R_{0}\rangle - 2e^{-2\kappa^{2}}|R_{g}\rangle.
\end{align}
Denoting by $|\psi_{0}^{\mathrm{aniso}}\rangle$ the ground state of the post-quench Hamiltonian and using Eq.~(\ref{eq:psi0}), we get
\begin{align}
     |\psi(t=0)\rangle = |\psi_{0}^{\mathrm{aniso}}\rangle - 2e^{-2\kappa^{2}}(1-e^{2\kappa^{2}A})|R_{g}\rangle.
\end{align}
Very close to the thin-cylinder limit, the graviton root state will be the correct $O(1)$ approximation to an eigenstate of the Hamiltonian, as confirmed by the numerics.
Thus, to first order, we can treat both $|\psi_{0}^{\mathrm{aniso}}\rangle$ and $|R_{g}\rangle$ as eigenstates and write the time-evolved state as
\begin{align}
    |\psi(t)\rangle &= |\psi_{0}^{\mathrm{aniso}}\rangle - 2e^{-2\kappa^{2}}(1-e^{2\kappa^{2}A})e^{-iE_{\gamma}t}\,|R_{g}\rangle,
\end{align}
with $E_\gamma$ being the energy of the graviton state.
Assuming that the combined anisotropy, coming from  the metric deformation and the stretching of the cylinder, is still small, $\kappa^{2}A \ll 1$, we can rewrite the above expression
\begin{align}
    |\psi(t)\rangle &\approx |R_{0}\rangle - 2e^{-2\kappa^{2}}(1 + 2\kappa^{2}A(1-e^{-iE_{\gamma}t}))|R_{g}\rangle.
\end{align}
The expression in the bracket can be rewritten  
\begin{align}
1 + 2\kappa^{2}A(1-e^{-iE_{\gamma}t}) \approx e^{2\kappa^2 A (1-e^{-iE_\gamma t})}.
\end{align}
Substituting into the previous equation,
\begin{align}
    |\psi(t)\rangle &\approx |R_{0}\rangle - 2\exp\left(-2\kappa^{2}[1 - A (1-e^{-iE_{\gamma}t})]\right)|R_{g}\rangle.
\end{align}
We recognize that this is of the same form as Eq.~(\ref{eq:psi0}), as the expression in the square bracket can be written as $1-\tilde{Q}\exp(i\tilde{\phi})$, with
\begin{eqnarray}\label{eq:bimetric}
 \tilde{Q}(t) = 2A \sin (E_{\gamma}t/2), \quad \tilde{\phi}(t) = \pi/2-E_{\gamma}t/2. \;\;\; \;\; 
\end{eqnarray}
These are nothing but the equations of motion of the linearized bimetric theory~\cite{Liu18}.  Thus, we have reproduced the graviton dynamics, which in the thin-cylinder limit reduces to the above two-level system dynamics. \cref{fig:fredkin geometric quench} confirms the existence of very regular metric oscillations at $L_2{=}3.6\ell_B$ and their agreement with the analytical expression in Eq.~(\ref{eq:bimetric}). From \cref{eq:graviton root state} we deduce that the energy of the graviton in the thin-cylinder limit is $E_{\gamma}= 2V_{023320} = 72\, e^{-14\kappa^{2}/3}\,$, which agrees with the frequency of the oscillations seen in Fig.~\ref{fig:fredkin geometric quench}.

\begin{figure}[tb]
\includegraphics[width=0.96\columnwidth]{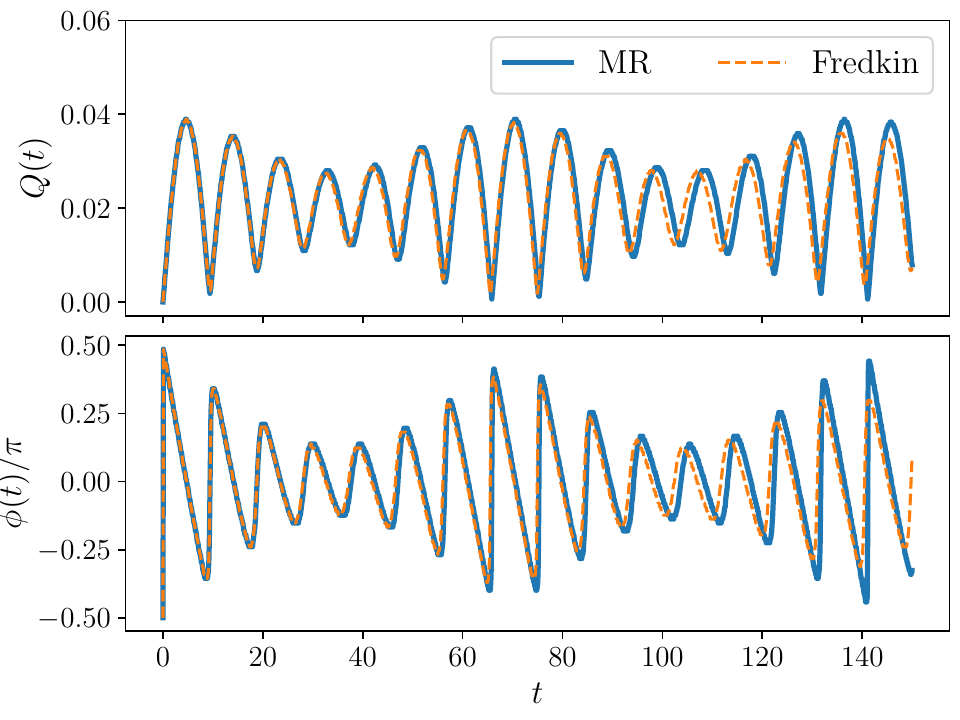}
	\caption{Comparison of the geometric quench dynamics between the Fredkin and full model at the cylinder circumference $L_2 {=} 5\ell_{B}$. The system size is $N_{e}{=}8$, $N_{\phi}{=}14$. The system is initialized in the isotropic ground state and then time-evolved by an anisotropic Hamiltonian with $Q {=} 0.02$. In this case, the dynamics is beyond the simple two-level system dynamics described by linearized bimetric theory, as the graviton does not correspond to a single eigenstate. The contribution of additional eigenstates to the spectral function gives rise to the beating pattern seen here. Nevertheless, there is still good agreement between the Fredkin and full model.}
    \label{fig:beats geometric quench}
\end{figure}

Notably, our Fredkin model still accurately captures the dynamics beyond the regime where it can be analytically treated as a two-level system. For example, around circumference $L_2 {\sim} 5\ell_{B}$, the graviton peak splits into a few smaller peaks close in energy. The resulting metric oscillations can be seen in \cref{fig:beats geometric quench}. There is now a slowly varying envelope that cannot be accounted for within the simple linearized bimetric theory in Eqs.~(\ref{eq:bimetric}).  At even larger circumferences, the structure of the graviton state becomes increasingly complicated, as there are many types of quadrupolar configurations of the root state. The spectrum undergoes dramatic transformations at intermediate values of $L_2$, as the hierarchy of energy scales in the Hamiltonian changes.  It is expected that close to the 2D limit and in the thermodynamic limit, the energy of the graviton stabilises, as the energy hierarchy stabilises too when $\kappa$ is small.

\section{Conclusions and discussion}\label{sec:conclusions}

We have formulated a one-dimensional qubit model that captures the MR state and its out-of-equilibrium properties on sufficiently thin cylinders with circumferences $L_2 {\lesssim} 7\ell_B$.  This was demonstrated by computing the overlap with the MR wave function and scaling of entanglement entropy with the size of the subsystem, as well as the dynamics following a geometric quench. One advantage of the proposed model is that its ground state can be written down exactly and it is amenable to efficient preparation on the existing quantum hardware, as we have demonstrated using the IBM quantum processor. At the expense of noise-aware error mitigation schemes~\cite{Kirmani2021}, these results can naturally be extended to probe the dynamics of the MR phase on quantum computers. This would also require an efficient optimally decomposed circuit to emulate trotterized evolution with our Hamiltonian~\eqref{eq:H_Pauli}, which is left to future work. 

There are some notable differences between the model presented here and previous studies of the $\nu{=}1/3$ Laughlin case~\cite{Kirmani2021}. While in the latter case, the truncated model can be easily adapted to either open or periodic boundary condition, in our case the torus boundary condition leads to considerable complications. For example, the hopping term in Eq.~(\ref{eq:mr term 3}) should no longer be neglected as it can act on the root state $1010\cdots 1010$, which is one of the Tao-Thouless torus ground states. Keeping this hopping term, in combination with Eq.~(\ref{eq:mr term 4}) we considered above, leads to more complicated models, none of which appears to be  frustration-free (i.e., until we exhaust all the terms in the Hamiltonian for a given finite system size). For this reason, our truncated model in Eq.~(\ref{eq:cylinder model}) applies primarily to the cylinder geometry. 

The previously mentioned caveats of boundary conditions highlight the fact that defining a truncated model is a nontrivial task.  Unlike the Laughlin case, where the Hamiltonian can be naturally truncated according to the magnitude of the matrix elements, leading to a frustration-free model, such a truncation scheme was not possible for the MR case.  The requirement of a frustration-free truncated Hamiltonian involves judiciously neglecting certain terms, which necessitates an independent demonstration of the model's validity.  In fact, similar difficulties are encountered even in the Laughlin case in going beyond the first-order truncation in Ref.~\cite{Nakamura2012} (see Appendix~\ref{app:motzkin}), and they become progressively more severe in higher members of the Read-Rezayi sequence~\cite{Papic14}. It would be useful if  a systematic approach could be developed for generating frustration-free models in all these examples, which would allow to controllably approach the isotropic 2D limit. The frustration-free property of the truncated model at $\nu{=}1/3$ has recently allowed to rigorously prove the existence of a spectral gap in that case~\cite{Nachtergaele2021,Nachtergaele2021_2,Warzel2022}. It would be interesting to see if such an approach could be generalized to the MR state and, potentially, to longer-range truncations.

As mentioned in Introduction, a unique feature of the MR state is the neutral fermion collective mode and the emergent supersymmetry relating that mode with the GMP mode we discussed above. It would be worth investigating  signatures of the neutral fermion mode in the proposed Fredkin model or other appropriate truncations of the MR Hamiltonian near the thin-cylinder limit. Unlike the GMP mode, which can be directly probed using the geometric quench, it is not known how to excite the neutral fermion mode. This is because the latter carries angular momentum $L{=}3/2$ in the long-wavelength limit. Therefore, it does not couple to the anisotropic deformations of the FQH fluid studied above. We leave the investigation of such dynamical probes and their implementation on quantum hardware to future work. 

{\sl Note added.} During the completion of this work, we became aware of a work by Causer \emph{et al.}~\cite{Causer2023} which finds evidence of anomalous thermalization dynamics and quantum many-body scars in a similar type of deformed Fredkin model. However, the model studied by Causer \emph{et al.}~\cite{Causer2022} assumes a different parameter range $|\tau|>1$, which is unphysical from the point of view of FQH realization considered here.

\begin{acknowledgments}
We would like to thank Zhao Liu and Ajit C. Balram for useful discussions.  
This work was supported by the Leverhulme Trust Research Leadership Award RL-2019-015.  This research was supported in part by the National Science Foundation under Grant No. NSF PHY-1748958. PG acknowledges support from NSF DMR-2130544 and infrastructural support from NSF HRD-2112550 (NSF CREST Center IDEALS). PG and AK acknowledge support from NSF DMR-2037996. AR acknowledges support from NSF DMR-2038028 and NSF DMR-1945395. We acknowledge the use of IBM Quantum services. We also thank the Brookhaven National Laboratory for providing access to IBM devices.
\end{acknowledgments}

\appendix

\section{Anisotropic interaction matrix elements for the Moore-Read state}\label{anisotropic MR}

We sketch the derivation of the interaction matrix elements when an anisotropic band mass is introduced. The interaction Hamiltonian can be written as:
\begin{equation}
    \hat{H} = \frac{1}{N_{\phi}} \sum_{\mathbf{p},\mathbf{q}} \bar{V}(\mathbf{p},\mathbf{q},-\mathbf{p}-\mathbf{q}) \, :\bar{\rho}(\mathbf{p}) \, \bar{\rho}(\mathbf{q}) \, \bar{\rho}(-\mathbf{p} - \mathbf{q}):
\end{equation}
where $\bar{\rho}(\mathbf{q}) = e^{-i q_{x} q_{y}/2} \, \sum_{j} e^{i q_{x} \kappa j} \, c^{\dagger}_{j+q_{y}/\kappa} c_{j}$ is the projected density operator, and $\bar{V}(\mathbf{p},\mathbf{q},-\mathbf{p}-\mathbf{q})$ is the interaction potential multiplied by the corresponding form factor:
\begin{equation}
     \bar{V}(\mathbf{p},\mathbf{q},-\mathbf{p}-\mathbf{q}) = F(\mathbf{p},\mathbf{q},-\mathbf{p}-\mathbf{q}) \, v(\mathbf{p},\mathbf{q},-\mathbf{p}-\mathbf{q}),
\end{equation}\\
where the form factor is
\begin{equation}
     F(\mathbf{p},\mathbf{q},-\mathbf{p}-\mathbf{q}) =  e^{-\mathbf{p}^{2}/4 - \mathbf{q}^{2}/4 - (\mathbf{p}+\mathbf{q})^{2}/4},
\end{equation}
and the interaction potential
\begin{align}
     v(\mathbf{p},\mathbf{q},&-\mathbf{p}-\mathbf{q}) =  \mathbf{p}^{4}\mathbf{q}^{2} + \mathbf{q}^{4}\mathbf{p}^{2} + \mathbf{q}^{4}(\mathbf{q}+\mathbf{p})^{2} + \nonumber \\
     & (\mathbf{q}+\mathbf{p})^{4}\mathbf{q}^{2} + \mathbf{p}^{4}(\mathbf{q}+\mathbf{p})^{2} + 
     (\mathbf{q}+\mathbf{p})^{4}\mathbf{p}^{2} 
\end{align}
is the Fourier transform of Eq.~(\ref{eq:MRHam}).

Anisotropic band mass tensor affects the single-electron wave functions -- see, e.g., Ref.~\cite{Yang12c}. Thus, it also modifies the matrix elements:
\begin{align}
    V_{j_{1}j_{2}j_{3}j_{4}j_{5}j_{6}}& \propto P_{g}(\{j_i\}) \  \mathrm{exp} \biggl( -\frac{\kappa^{2}}{2g_{11}}(\sum j^{2}_{i} - \frac{1}{6}(\sum j_{i})^{2}) \nonumber \\ &+ \frac{i\kappa^{2} g_{12}}{2 g_{11}} (j^{2}_{6}+j^{2}_{5}+j^{2}_{4} - j^{2}_{3} - j^{2}_{2} - j^{2}_{1}) \biggr)
\end{align}
Just as in the isotropic case, the polynomial $P_{g}$ is tightly constrained: it has to be antisymmetric in the pairs $(j_{1},j_{2})$, $(j_{1},j_{3})$, $(j_{2},j_{3})$, $(j_{4},j_{5})$, $(j_{5},j_{6})$, $(j_{5},j_{6})$, and its maximum total degree is $6$. The only such polynomial is the one that appears in the isotropic case. Therefore the only contribution of the metric in the prefactor is a constant. The final form will be:
\begin{align}
    V_{j_{1} \dots j_{6}}& \propto \frac{\kappa^{8}}{g_{11}^{4}} (j_{1}-j_{2})(j_{1}-j_{3})(j_{2}-j_{3})(j_{6}-j_{4}) \nonumber \\ & (j_{6}-j_{5})(j_{5}-j_{4}) \  \mathrm{exp} \biggl( -\frac{\kappa^{2}}{2g_{11}}(\sum j^{2}_{i} - \frac{1}{6}(\sum j_{i})^{2}) \nonumber \\ &+ \frac{i\kappa^{2} g_{12}}{2 g_{11}} (j^{2}_{6}+j^{2}_{5}+j^{2}_{4} - j^{2}_{3} - j^{2}_{2} - j^{2}_{1}) \biggr).
\end{align}

\section{Nonlocal string order in the Fredkin chain}

The nonlocal constraint that defines Dyck paths hints that the Fredkin ground state might have interesting behavior in certain nonlocal order parameters. This is reinforced by the fact that such nonlocal correlations were found in the spin-1 analog of our model, the deformed Motzkin chain~\cite{Barbiero17}. The natural correlations to probe in the Fredkin chain are the string orders discussed in Ref.~\cite{Kim20} in connection to spin-$1/2$ ladders and the Majumdar-Ghosh chain:
\begin{equation} \label{eq:string order}
    O_{\mathrm{even/odd}} = \lim_{|i-j|\to\infty} \biggl< \big(S^{z}_{i} + S^{z}_{i+1}\big)  e^{i\pi\sum_{l=i+2}^{j-1}S^{z}_{l}} \bigl(S^{z}_{j} + S^{z}_{j+1} \bigr) \biggr>
\end{equation}
where for $O_{\mathrm{even/odd}}$ the sites $i,j$ are both even/odd, respectively.

\begin{figure}[htb]
    \centering
    \includegraphics[width=1\linewidth]{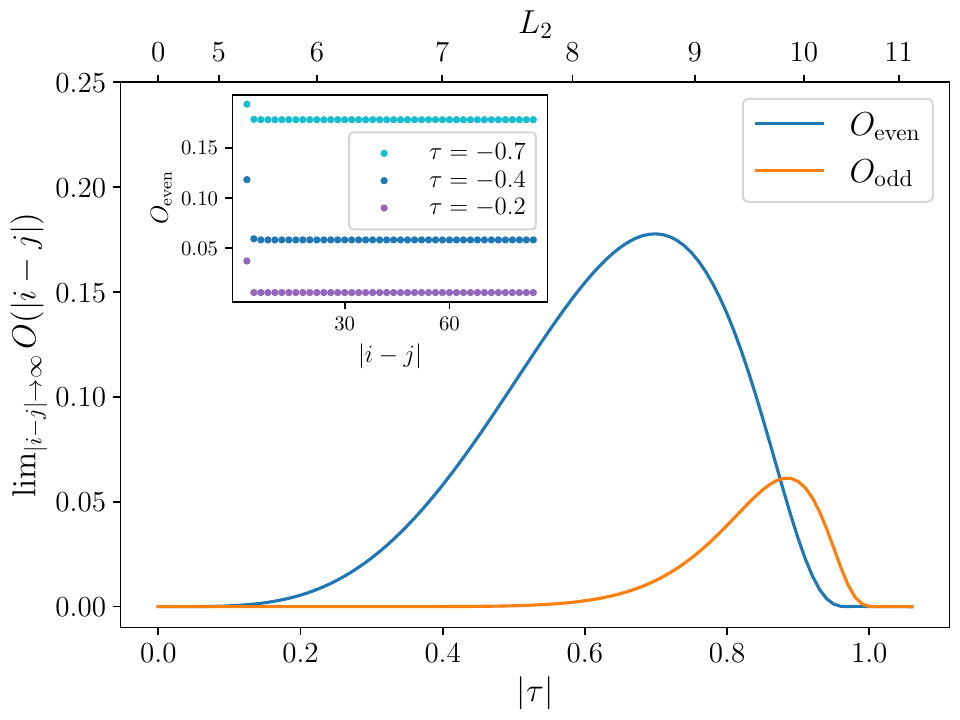}
    \caption{The behavior of $O_{\mathrm{even/odd}}$ as a function of the deformation parameter $\tau$. String order is not present at $\tau=0$, i.e., when the ground state is a product state. For $|\tau|<1$ the string order parameter increases but drops quickly at $|\tau|=1$ where the gap closes. The difference in magnitude between $O_{\mathrm{even}}$ and $O_{\mathrm{odd}}$ is a result of stronger bonds that form between sites $2i$ and $2i+1$, such that all spins are paired up. The inset shows the behavior of $O_{\mathrm{even}}$ as a function of $|i-j|$, demonstrating that the nonlocal correlations quickly stabilize to a $\tau$-dependent value. All numerical results are obtained from a chain with $N=100$ spins, where the values are already converged.
    }
    \label{fig:string operators}
\end{figure}

Using the MPS representation (\ref{eq:MPS deformed}), we test the Fredkin ground state for nonlocal order, shown in \cref{fig:string operators}. First, note that nondecaying expectation values are only found inside the $|\tau|<1$ phase (as the inset shows), whereas in the $|\tau| >1$ ``domain-wall" phase these nonlocal correlations decay. This suggests that in the $|\tau|<1$ ``antiferromagnetic" phase, short range valence bonds form between consecutive spins. We also notice that generally $O_{\mathrm{even}}$ is higher in magnitude compared to $O_{\mathrm{odd}}$. Given that the spin chain always has even length, the favored arrangement is where all spins are paired (i.e. $(0,1), (2,3),\dots,(N-2,N-1)$), as opposed to the case where the first and the last spins remain unpaired. This implies the bonds starting on an even index will be stronger.

\section{Motzkin chain as an effective truncated model of the $\nu{=}1/3$ Laughlin state} \label{app:motzkin}

In this Appendix, for the sake of completeness, we show that the Motzkin chain~\cite{Bravyi12,Movassagh16,Movassagh17,ZhangZhao17} -- a closely related spin-1 cousin of the Fredkin chain -- captures the properties of the $\nu{=}1/3$ Laughlin state. This model contains more terms compared to the model derived in Ref.~\cite{Nakamura2012} and hence captures the physics of the Laughlin state over a slightly larger range of cylinder circumferences. 

The model in Ref.~\cite{Nakamura2012} can be derived via a similar method to the one presented above, but for a 2-body interaction given in terms of $V_1$ Haldane pseudopotential~\cite{Prange87}. The corresponding matrix elements are now given by
\begin{eqnarray}
  V_{j_1j_2j_3j_4} = (j_1-j_2) (j_4-j_3)e^{-\frac{\kappa^2}{4}\left[ (j_1-j_2)^2 + (j_3-j_4)^2 \right]}.  \;\;\;\;\;
\end{eqnarray}
The minimal model beyond the extreme thin-cylinder limit from Ref.~\cite{Nakamura2012} can be written in the positive-semidefinite form
\begin{equation}
    H'_{\mathrm{L}} = \sum_{i} \left( Q^{\dagger}_{i}Q_{i} + P^{\dagger}_{i}P_{i} \right)
\end{equation}
where 
\begin{equation}
    Q_{i} = \alpha_{i} c_{i+1}c_{i+2} + \gamma_{i}c_{i}c_{i+3} \,, \quad P_{i} = \beta_{i}c_{i}c_{i+2}
\end{equation} 
and
\begin{equation}
    \alpha = \sqrt{V_{0110}}, \quad \beta=\sqrt{V_{0220}}, \quad \gamma = e^{2i\kappa^{2}\frac{g_{12}}{g_{11}}}\sqrt{V_{0330}}
\end{equation}

The only configurations which are dynamically connected to the root state are those that can be obtained from applying squeezing operators $\hat{S}_{i} = c_{i+1}^{\dagger}c_{i+2}^{\dagger}c_{i+3}c_{i}$ to $100100 \ldots 001$. This connected component of the Hilbert space can be mapped to a spin-$1$ model by considering unit cells of three magnetic orbitals, whose occupations can only take the following patterns:
\begin{equation}
    |010\rangle \to |\mathrm{o}\rangle\,, \quad
    |001\rangle \to |+\rangle\,, \quad
    |100\rangle \to |-\rangle.
\end{equation}
Thus, we can write the model of Ref.~\cite{Nakamura2012} as 
\begin{equation}\label{eq:BN model}
    H'_{\mathrm{L}} = \sum_{i=0}^{N-2} \mathcal{P}^{\varphi_{\mathrm{L}}(v)}_{i,i+1}
\end{equation}
where $|\varphi_{\mathrm{L}}(v)\rangle = |{+-} \rangle - v |\mathrm{oo}\rangle$ and $v= -\sqrt{V_{0330}/V_{0110}} = -3 \exp(-2\kappa^{2})$. It is important to notice there are no boundary conditions -- they are not necessary if the mapped Hilbert space is used (which entails constraints, e.g. configurations with  the first spin $|-\rangle$ and the last spin $|+\rangle$ being disallowed). 

\subsection{Extension to the Motzkin chain}

A natural attempt to improve the model in \cref{eq:BN model} would be to extend the truncation, $P_{i} \to \beta_{i}c_{i}c_{i+2} + \delta_{i}c_{i}c_{i+4}$. The newly obtained Hamiltonian $H^\prime$ would have the following off-diagonal actions:
\begin{align}
    H''_{\mathrm{L}} | \dots 100 \, 010 \dots \rangle &= \beta \delta | \dots 010 \, 100 \dots \rangle \nonumber \\
    H''_{\mathrm{L}} | \dots 010 \, 001 \dots \rangle &= \beta \delta | \dots 001 \, 010 \dots \rangle 
\end{align}
and the Hermitian conjugates. In the spin-$1$ mapping, these mean
\begin{align}
    H''_{\mathrm{L}} | \dots {-} \, \mathrm{o} \dots \rangle &= \beta \delta | \dots \mathrm{o} \, {-} \dots \rangle \nonumber \\
    H''_{\mathrm{L}} | \dots \mathrm{o} \, {+} \dots \rangle &= \beta \delta | \dots {+} \, \mathrm{o} \dots \rangle 
\end{align}
However, the fermionic Hamiltonian also produces hoppigns of the follwoing kind:
\begin{align}
    H''_{\mathrm{L}} | \dots 100 \, 100 \dots \rangle &= \beta \delta | \dots 011 \, 000 \dots \rangle \nonumber \\
    H''_{\mathrm{L}} | \dots 001 \, 001 \dots \rangle &= \beta \delta | \dots 000 \, 110 \dots \rangle 
\end{align}
These break our spin mapping, and connect the entire Hilbert space. Even though we only keep 2 types of off-diagonal terms, we no longer obtain a significant reduction in complexity, and in fact we find numerically that the zero-mode property is also lost. Thus, we focus only on the spin model instead. The extension of the Hamiltonian in \cref{eq:BN model} therefore takes the form:
\begin{equation}\label{eq:Motzkin model}
    H_{\mathrm{M}} = \sum_{i=0}^{N-2} \mathcal{P}^{\varphi_{\mathrm{L}}(v)}_{i,i+1} + \mathcal{P}^{U(w)}_{i,i+1} + \mathcal{P}^{D(w)}_{i,i+1},
\end{equation}
where we introduced the projectors on the states $|U(w)\rangle = |{+}\mathrm{o}\rangle - w |\mathrm{o}+\rangle$ and $|U(w)\rangle = |\mathrm{o}-\rangle - w |{-}\mathrm{o}\rangle$, where $w = -\sqrt{V_{0440}/V_{0220}} = -2 \exp (-3\kappa^{2})$. These implement the additional the additional terms in our truncation, while keeping the spin Hamiltonian $2$-local.

\begin{figure}[t]
    \centering
    \includegraphics[width=\linewidth]{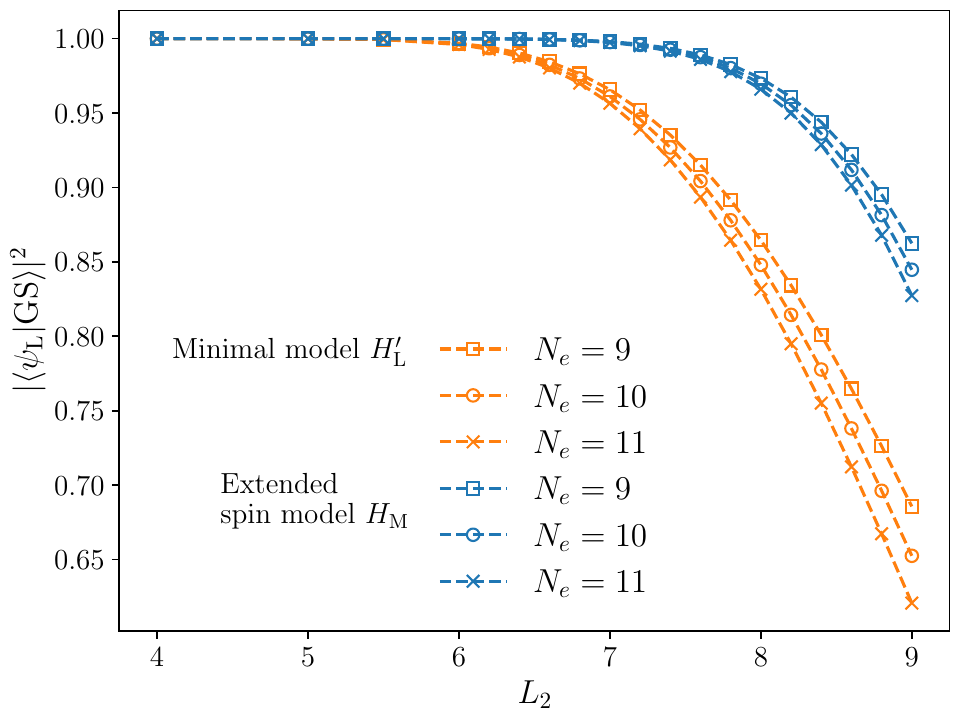}
    \caption{Squared overlaps of the ground states of models in \cref{eq:BN model} (the minimal model $H'_{\mathrm{L}}$) and \cref{eq:Motzkin model} (the extended spin model $H_{\mathrm{M}}$) with the ground state of the untruncated $V_{1}$ Hamiltonian. The extended spin model captures the properties of the Laughlin state up to cylinder circumferences of $L_{2} \approx 8\,l_{B}$, where the overlaps are $\gtrsim 95 \%$.}
    \label{fig: motzkin overlaps}
\end{figure}

The Hamiltonian in \cref{eq:Motzkin model} represents a particular deformation of the Motzkin spin chain introduced in \cite{ZhangZhao17}. It has a unique, zero-energy ground state which is equal to an area-weighted sum of Motzkin paths, $p \in \mathcal{M}_{N}$: 
\begin{equation} \label{eq:motzkin gs}
    |\psi_{0}^{\mathrm{M}} \rangle = \mathcal{N}^{-1} \sum_{p \in \mathcal{M}_{N}} v^{A_{\square}(p)} w^{A_{\triangle}(p)} |p\rangle.
\end{equation}

\begin{figure}[ht]
    \centering
    \includegraphics[width=\linewidth]{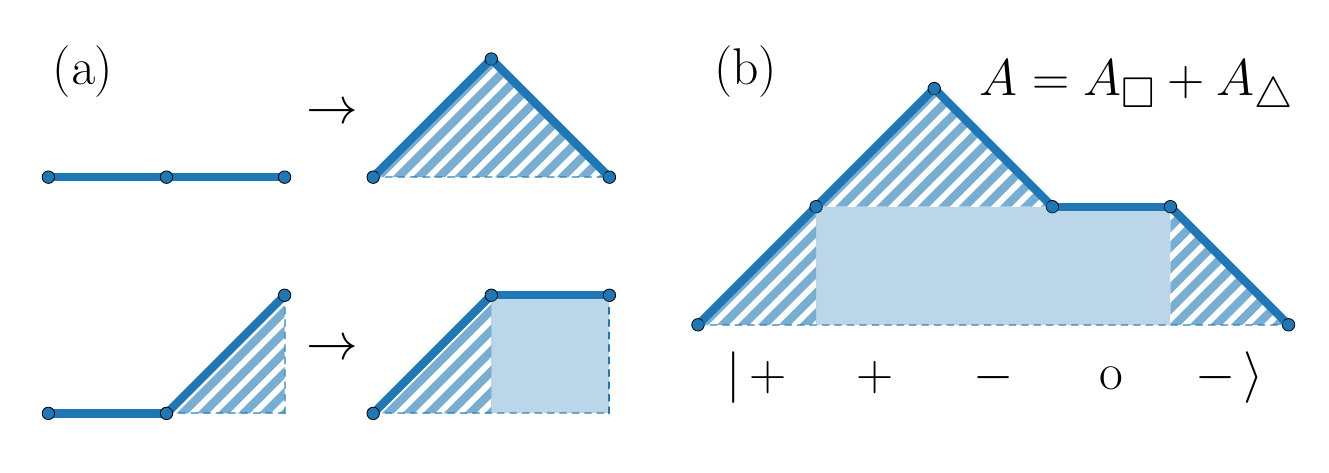}
    \caption{(a): Two types of allowed moves in the Motzkin chain. The upper move corresponds to $|\mathrm{oo}\rangle \to |{+-}\rangle$, while the bottom one shows $|\mathrm{o}{+}\rangle \to |{+}\mathrm{o}\rangle$ ($|{-}\mathrm{o}\rangle \to |\mathrm{o}{-}\rangle$ is omitted for brevity). Although each step increases the area of the path by the same amount, the corresponding weight in \cref{eq:motzkin gs} scales differently depending on the move.
    (b): One type of allowed configuration in the ground state, that was not present in the model \cref{eq:BN model}. The sketch shows how the total area is divided into $A_{\square}$ and $A_{\triangle}$, determining its weight in the ground state.}
    \label{fig:motzkin moves}
\end{figure}

\cref{fig: motzkin overlaps} shows that this ground state has good overlap with the Laughlin over a larger range of circumferences. Notice that in the Motzkin ground state with $v=w$, the weights of a path $p$ are $v^{A(p)}$, i.e. only dependent on the total area and not its shape. \cref{fig:motzkin moves} illustrates how this is different from \cref{eq:motzkin gs}.

Similar to the Fredkin chain discussed in \cref{app:mps fredkin}, the ground state of the Motzkin chain also has an exact MPS representation in terms of matrices:
\begin{equation}
    A^{\mathrm{o}}_{jk} = w^{j-1} \delta_{j,k} \quad A^{+}_{jk} = v^{1/2} w^{j-1} \delta_{j+1,k} \quad A^{-} = (A^{+})^{T}
\end{equation}
The Motzkin chain with equal deformation parameters $v=w$ has been studied in depth in the literature. Its gap for $v<1$ has been proven \cite{Andrei22}, and based on our analogy with the Laughlin state we conjecture that the gap survives for $w \leq v$.

\subsection{Laughlin graviton root state and geometric quench dynamics} \label{app:graviton laughlin}

Here we analyze the graviton root state and dynamics following the geometric quench for the $\nu{=}1/3$ Laughlin state, following a similar approach the MR state in Sec.~\ref{sec:dynamics}. As explained in the main text, we can use the SMA ansatz~\cite{Girvin85,Girvin86} to identify the GMP state with nonzero momentum $\mathbf{k}$:
\begin{equation}
    |\phi_{\mathbf{k}}\rangle = \overline{\rho}_{\mathbf{k}} |\psi_{0}\rangle = e^{\frac{ik_{x}k_{y}}{2}}\sum_{j} e^{ik_{x}\kappa j} c^{\dagger}_{j+k_{y}/\kappa}c_{j} |\psi_{0}\rangle
\end{equation}
In the thin-torus limit, the ground state is the root state $|R_0\rangle = |1001001\dots\rangle$. Thus, the graviton root state with momentum $\mathbf{k} = 2\kappa \hat{y}$
 is given by
 \begin{equation}
    |R_g^{(2)}\rangle \propto |1100001001\dots\rangle + |1001100001\dots\rangle + \dots
\end{equation}
In the extreme thin-cylinder limit, these states are degenerate in energy, and the first product state is the Jack root state \cite{BoYang12}, from which all states that follow can be obtained by applying a sequence of squeezes. 

\begin{figure}[tb]
    \includegraphics[width=1\columnwidth]{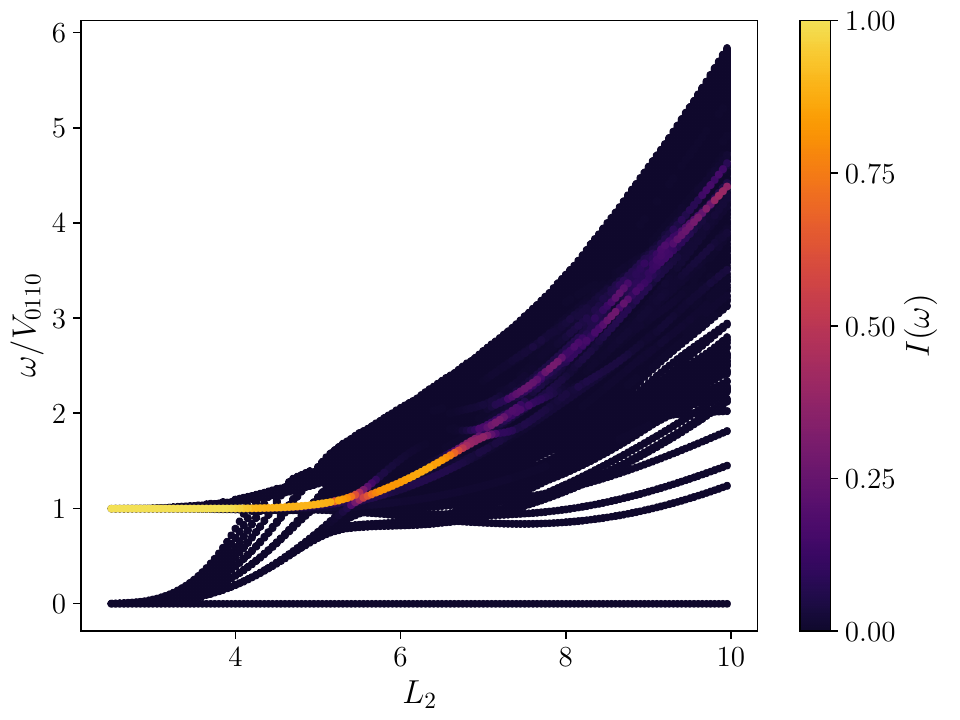}
	\caption{The spectral function $I(\omega)$ for the untruncated Laughlin Hamiltonian at $\nu{=}1/3$ as the cylinder circumference $L_2$ is varied between the isotropic 2D limit and the thin-cylinder limit. The system size is $N=9$ electrons, $N_\phi=25$.}
	\label{fig: laughlin spectral fn n=7 full}
\end{figure}

However, as the geometric quench preserves momentum, to identify the long-wavelength limit of the graviton state we rely on spectral function $I(w)$ from Eq.~(\ref{eq:spectral fn}). In the present case, $\hat O$ is a 2-body operator~\cite{Liou19} with matrix elements 
\begin{align}\label{eq:2bodyquadrupole}
    O_{j_{1}j_{2}j_{3}j_{4}} &=  \delta_{j_{1}+j_{2},j_{3}+j_{4}} (j_{1}-j_{2})(j_{3}-j_{4}) \nonumber \\
    &\times \left(\sum j_{i}^{2} - \frac{1}{4}(\sum j_{i})^{2}\right) \nonumber \\
    &\times \exp \left[ -\frac{\kappa^{2}}{2}\left(\sum j^{2}_{i} - \frac{1}{4}(\sum j_{i})^{2}\right) \right].
\end{align}
The spectral function $I(\omega)$ for the $\nu{=}1/3$ Laughlin state is plotted in Fig.~\ref{fig: laughlin spectral fn n=7 full}. Similar to the MR case in Fig.~\ref{fig:spectral fn fredkin}, we see that the graviton undergoes a nontrivial evolution as the cylinder circumference is varied, with clear avoided crossings in the evolution. In the thin-cylinder limit, the gap of the graviton can be accurately estimated from the dominant matrix element in the Hamiltonian. 

\begin{figure}[b]
    \includegraphics[width=1\columnwidth]{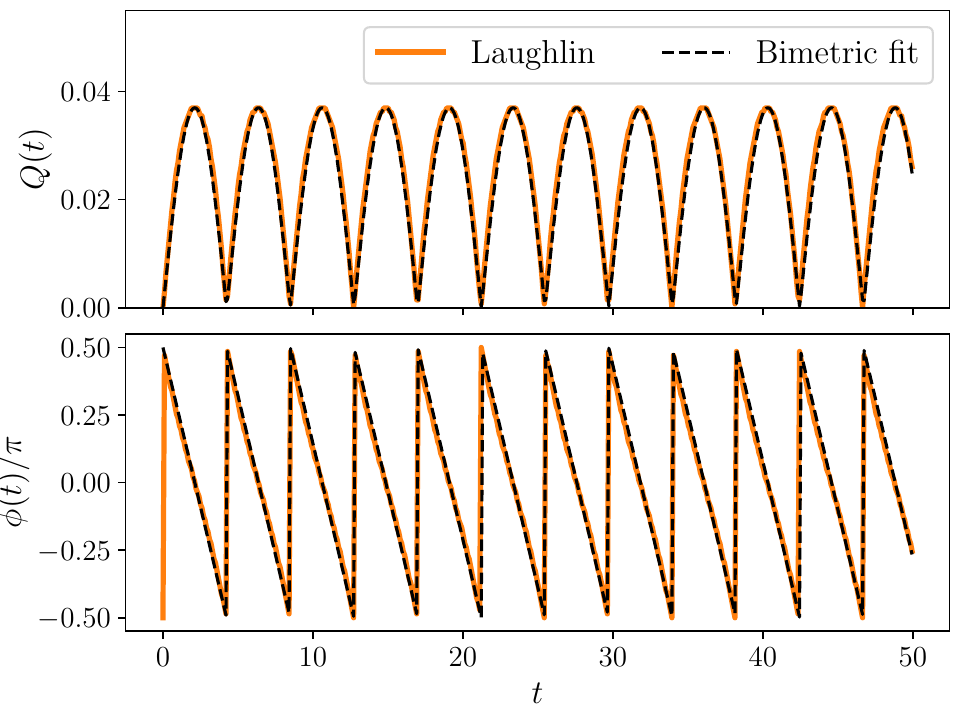}
	\caption{Geometric quench dynamics in the $\nu{=}1/3$ Laughlin state.  The system size is $N_{e}{=}6$, $N_{\phi}{=}16$, and the circumference is $L_2 = 2.8\ell_{B}$. The system is initialised in the isotropic ground state and then time-evolved by the anisotropic Hamiltonian with $Q{=}0.02$. The resulting dynamics is in excellent agreement with the linearized bimetric theory, shown by dashed lines.}
	\label{fig: metric quench BN}
\end{figure}

The graviton state is given by acting on the ground state with the quadrupole operator (\ref{eq:2bodyquadrupole}). 
From the model in \cref{eq:BN model}, we also know that the ground state is approximated by
\begin{align}
    |\psi_{0}\rangle &= \prod_{i}\big(1-\sqrt{V_{0330}/V_{0110}} \,e^{2i\kappa^{2}g_{12}/g_{11}} \hat{S}_{i} \big) |R_0\rangle \nonumber \\
    & \approx |R_0\rangle - 3e^{-2\kappa^{2}\frac{1-ig_{12}}{g_{11}}}\sum_{i} \hat{S}_{i}|R_0\rangle \nonumber \\
    &\approx |R_0\rangle - 3\exp\big[ -2\kappa^{2}\big(1-Qe^{i\phi} \big) \big] \sum_{i} \hat{S}_{i}|R_0\rangle ,
\end{align}
where we assumed $e^{-2\kappa^{2}}$ and the metric anisotropy $Q,\phi$ to be small. The graviton is then approximated by:
\begin{align}
    |\psi_{g}\rangle &= \hat{O} |\psi_{0}\rangle \propto e^{-\frac{5\kappa^{2}}{2}}\bigg(\sum_{i} \hat{S}_{i}|\psi_{0}\rangle + \mathcal{O}\big(e^{-2\kappa^{2}}\big)\bigg).
\end{align}
From here we deduce the graviton root state, 
\begin{equation}
    |\psi_{g} \rangle = \sum_{i} \hat{S}_{i}|\psi_{0}\rangle = |01100010\dots\rangle + |10001100\dots\rangle + \dots
\end{equation}
Similar to the MR case, the graviton root state here is also proportional to the first order squeezes and it encodes the simplest quadrupole structure of the form $-++-$ in each unit cell.

Repeating the same steps as in Eqs.~(\ref{eq:timeevolvedpsi})-(\ref{eq:bimetric}) of the main text, from the graviton root state we can determine the time-evolved state, showing that it takes the form (at first order) 
\begin{align}
    |\psi(t)\rangle \approx |R_0 \rangle - 3e^{-2\kappa^{2}}(1 + 2\kappa^{2}A(1-e^{-iE_{\gamma}t}))|R_g\rangle,
\end{align}
which has the identical form to the linearized bimetric theory in Eq.~(\ref{eq:bimetric}). This agreement is confirmed in Fig.~\ref{fig: metric quench BN}. 

\bibliography{biblio_fqhe}

\end{document}